\documentclass[a4paper,12pt]{article}
\usepackage{amsmath, amssymb, authblk, a4wide, bm, cancel, color, graphicx, enumerate, appendix, youngtab, cite, hyperref, verbatim, slashed, subcaption, float, bm}
\usepackage[utf8]{inputenc}
\usepackage{amsfonts}
\usepackage{mathtools}
\newcommand*{\diffdchar}{\mathrm{d}}
\newcommand*{\dd}{\mathop{\diffdchar\!}}
\usepackage{ifpdf}
\graphicspath{{./img/}}
\usepackage{hyperref}

\DeclarePairedDelimiter\abs{\lvert}{\rvert}%
\makeindex

\begin{document}
\title{\vspace{2cm}
  \Large{\textbf{Constraining fine tuning in composite Higgs models with partially composite leptons}}}
\author[1]{\small {\bf James Barnard}}
\author[2]{\small{\bf Daniel Murnane}\thanks{\texttt{daniel.murnane@adelaide.edu.au, ORCID: 0000-0003-4046-4822}}}
\author[2]{\small{\bf Martin White}\thanks{\texttt{martin.white@adelaide.edu.au}}}
\author[2]{\small{\bf Anthony G. Williams}\thanks{\texttt{anthony.williams@adelaide.edu.au, ORCID: 0000-0002-1472-1592}}}
\affil[1]{ARC Centre of Excellence for Particle Physics at the Terascale, School of Physics, University of Melbourne, Victoria 3000, Australia}
\affil[2]{ARC Centre of Excellence for Particle Physics at the Terascale, Department of Physics, University of Adelaide, South Australia 5005, Australia}
\date{}
\maketitle

\begin{abstract}
Minimal Composite Higgs Models (MCHM) have long provided a solution to the hierarchy problem of the Standard Model, yet suffer from various sources of fine tuning that are becoming increasingly problematic with the lack of new physics observations at the LHC. We develop a new fine tuning measure that accurately counts each contribution to fine tuning (single, double, triple, etc) that can occur in a theory with $n_p$ parameters, that must reproduce $n_o$ observables. We then use a novel scanning procedure to perform a comprehensive study of three different two-site, 4D, $SO(5)\rightarrow SO(4)$ MCHMs with all third generation fermions included, distinguished by the choice of the lepton embeddings. These are the MCHM$^{\textbf{5-5-5}}_{\textbf{5-5-5}}$, MCHM$^{\textbf{5-5-5}}_{\textbf{14-14-10}}$ and MCHM$^{\textbf{5-5-5}}_{\textbf{14-1-10}}$, where MCHM$^{q-t-b}_{l-\tau-\nu}$ has the lepton doublet partner in representation $l$, tau partner in representation $\tau$, and so on. We find that embedding at least one massive lepton in the symmetric \textbf{14} of $SO(5)$ reduces the tuning for the case of low top partner masses (in line with previous results), but that this is balanced against the increased complexity of the model when one properly accounts for all sources of fine tuning. We study both the current relative fine-tuning of each scenario, and the future prospects. Noting that the different scenarios behave differently with respect to future improvements in collider measurements, we find that the MCHM$^{\textbf{5-5-5}}_{\textbf{14-1-10}}$ enjoys a relatively low increase in fine tuning even for a future lower bound on the top partner masses of 3.4 TeV (or equivalently a maximum Higgs-fermion or Higgs-gluon coupling deviation of 2\%).
\end{abstract}

\newpage
\tableofcontents

\section{Introduction}

The hierarchy problem of the Standard Model (SM) has long been used to motivate the existence of new physics at the TeV scale. A well-known and compelling extension of the SM involves the replacement of the elementary Higgs boson by a composite state that emerges as a pseudo-Nambu-Goldstone boson of a new, spontaneously broken global symmetry \cite{kaplan1984,kaplan1984b,kaplan1985}. As seen in the case of the pion mass in QCD, such a theory naturally contains a hierarchy between a characteristic mass scale associated with some new fundamental physics (for a example a strongly-coupled new sector), near which heavy resonances are expected to congregate, and an anomalously light object. One can also use compositeness to explain the fermion mass hierarchies~\cite{kaplan1991, Gherghetta:2000qt}. An important innovation in more recent work is the notion of ``partial compositeness'', in which SM fermions and gauge fields are a mixture of elementary fields and the new composite states. Whilst the precise choice of global symmetry remains open (see \cite{contino2003,agashe2005,contino2007,giudice2007,pomarol2012,marzocca2012,pappadopulo2013,sannino2014} for examples), the model based on $SO(5) \times U(1)_X \rightarrow SO(4)\times U(1)_X$ is known as the Minimal Composite Higgs Model (MCHM), since this is the smallest symmetry group consistent with custodial symmetry that leads to exactly four Nambu-Goldstone bosons~\cite{contino2006,contino2007b}. This symmetry breaking pattern fixes the embedding of the SM gauge sector but leaves considerable freedom of choice in the embedding of the SM fermion sector, in which case it is most usual to explore several possible options. 

The current lack of evidence for BSM physics at the LHC is telling us that the compositeness scale in any realised composite Higgs scenario is probably significantly higher than the electroweak scale, in which case it is worth critically examining the level of fine tuning in these theories. Fine tuning comes from a variety of sources, the most obvious being that the Higgs vacuum expectation value (VEV) must be kept below the compositeness scale. One must also make the theory reproduce a variety of observations including the Higgs VEV, the Higgs mass and the fermion masses, each of which can in principle contribute an independent source of fine tuning. It is known, for example, that the Higgs mass is specifically correlated with the mass of the lightest top partner for certain choices of the embedding of the composite states in $SO(5)$~\cite{matsedonskyi2012,marzocca2012,pomarol2012,panico2012,pappadopulo2013,Barnard:2013hka,Han:2014qia}, making light top partners essential for naturalness in these models.

In MCHM theories without partially composite leptons, one observes that embedding the quarks in {\bf 5} or {\bf 10} representations of $SO(5)$ leads to a well-known ``double tuning'' effect, which can ultimately be traced back to the calculation of the Higgs potential. One must first tune the parameters to obtain a Higgs VEV below the compositeness scale. Even then, this does not lead to viable electroweak symmetry breaking if one includes only the leading contribution to the Higgs potential, since the resulting functional form (in the Higgs field) has a minimum at the origin. The inclusion of formally subleading contributions produces a second term which can give rise to electroweak symmetry breaking provided that the parameters that enter the coefficient of the leading term can be tuned sufficiently far down from their natural size, independently of the initial tuning.

Embedding at least one quark chirality in the {\bf 14} representation provides a partial solution to this problem, since one now obtains two different functional forms in the Higgs potential at leading order. In practise, however, a large ad-hoc tuning is then required to reduce the Higgs mass, which turns out to be generically much higher than the electroweak scale in these models.  It has recently been shown, however, that a more elegant solution is to include partially composite leptons in the model, since they must be accounted for in any complete description in any case, and they can introduce extra effects~\cite{carmona2015}. Naively, it might be expected that the leptons play only a small role in the phenomenology of the model. At any given order in the Higgs potential calculation, for example, the lepton contribution (assuming similar embeddings to the quarks) will be considerably smaller than the quark contribution due to the much smaller Yukawa couplings and, to a lesser extent, the lack of a colour factor. However, if one puts the leptons in a {\bf 14} representation, but the quarks in {\bf 5} or {\bf 10} representations, one obtains two functional forms in the Higgs potential \emph{at leading order} for the leptons, with a suppressed contribution from the quarks. The two terms in the Higgs potential naturally come out to be of similar order, with less need for an additional ad-hoc tuning compared to the case of the initial quark approach.

In previous work by a subset of the current authors~\cite{Barnard:2015ryq}, comprehensive scans of MCHM scenarios without partially composite leptons were performed, with three different choices of the SM top quark embedding (and lighter fermions neglected). In each case, the regions of the parameter space consistent with the Higgs VEV, top quark mass and the Higgs mass were identified and used to obtain current and projected constraints on fine tuning as a function of existing and hypothetical limits on the top partner masses, charged vector-boson resonance masses, Higgs coupling deviations and the compositeness scale. In this paper, we revisit this work with the addition of the partially composite third generation leptons, also introducing the possibility of partially composite $b$ quarks for good measure. As such, the matter content of the models considered includes composite fermions with the quantum numbers of the heaviest flavour of SM quarks and leptons. We consider one quark representation (since the details of the tuning will be relatively insensitive to the choice of {\bf 5} or {\bf 10} representations for the quarks), and scan over three different lepton representations. These are the composite lepton doublet, tau and tau neutrino in \textbf{5-5-5} respectively, in the symmetric-antisymmetric \textbf{14-14-10}, or in the fully-composite tau \textbf{14-1-10}. These models have between 19 and 25 input parameters, and we must find regions of this large parameter space that correctly reproduce the $b$ quark, $\tau$ lepton and top quark masses, in addition to the Higgs mass and VEV. Such a large number of observables and parameters requires a sophisticated treatment, both in the definition of the fine tuning measure, and the method used to find the small regions of the considerable parameter space volume that give rise to the correct SM behaviour. We approach the first problem by generalising the fine-tuning measure of~\cite{Barnard:2015ryq} to cope with $N$ observables, and in doing so we count the total fine tuning in a more accurate way than in ~\cite{carmona2015} which leads to interesting conclusions. We solve the second problem by using a combination of the {\tt MultiNest} implementation of the nested sampling algorithm, plus a second stage of MCMC sampling, to efficiently find the desired regions in our candidate MCHM parameter spaces. Such regions are punishingly hard to find by random means, and our approach will be useful in the study of other composite Higgs models, all of which can be expected to rely on delicate cancellations between parameters to produce known phenomenology.

The rest of our paper is structured as follows. We provide a short overview of the two-site 4D MCHM in Section \ref{model}, and review the standard derivations of the relevant fermion expressions from the composite fermion Lagrangian. This includes formulae for the SM observables as functions of the model parameters. Our scanning technique is described in Section \ref{scanning}. In Section \ref{tuning} we outline how to deal with the tuning of $N$ observables, and present a computationally effective method of producing a total fine tuning. Our results are presented in Section \ref{results}, before we present our final conclusions in Section~\ref{sec:conclusions}.

\section{The Minimal Composite Higgs}
\label{model}
\subsection{Model overview}
A particularly elegant mechanism for realising a composite Higgs boson is to take a composite sector that emerges from a confining gauge theory (e.g. the theories considered in~\cite{Miransky:1988xi,Galloway:2010bp,Caracciolo:2012je,Barnard:2013zea,Ferretti:2013kya,Marzocca:2013fza,Cacciapaglia:2014uja,Ferretti:2014qta,vonGersdorff:2015fta,Vecchi:2015fma}) and associate the Higgs with a pseudo Nambu-Goldstone boson (pNGB) that derives from the spontaneous breaking of a global symmetry. The symmetry breaking pattern $SO(5)\rightarrow SO(4)$ preserves precision electroweak measurements through compatibility with custodial symmetry~\cite{Agashe:2003zs}, and produces four pNGBs which is exactly the number required to form a Higgs doublet. The interactions of the pNGBs are determined by low energy theorems, and hence the only remaining task is to specify the form of the interactions between the SM fermions and the composite sector, which boils down to choosing the precise embeddings of the elementary fermions in representations of $SO(5)$.  In this paper, we focus on the two-site 4D models that were previously described in~\cite{de2012,panico2012,carena2014}. The mechanics of the collective breaking will be broadly summarised below. However, this study uses well-established bosonic expressions, e.g., derived in \cite{Barnard:2015ryq,carena2014}, and we refer the reader to these for a pedagogical guide. We will delve into more detail of the behaviour of the fermion sector under the breaking. 

\begin{figure}[h]
\centering
\includegraphics[scale=0.5]{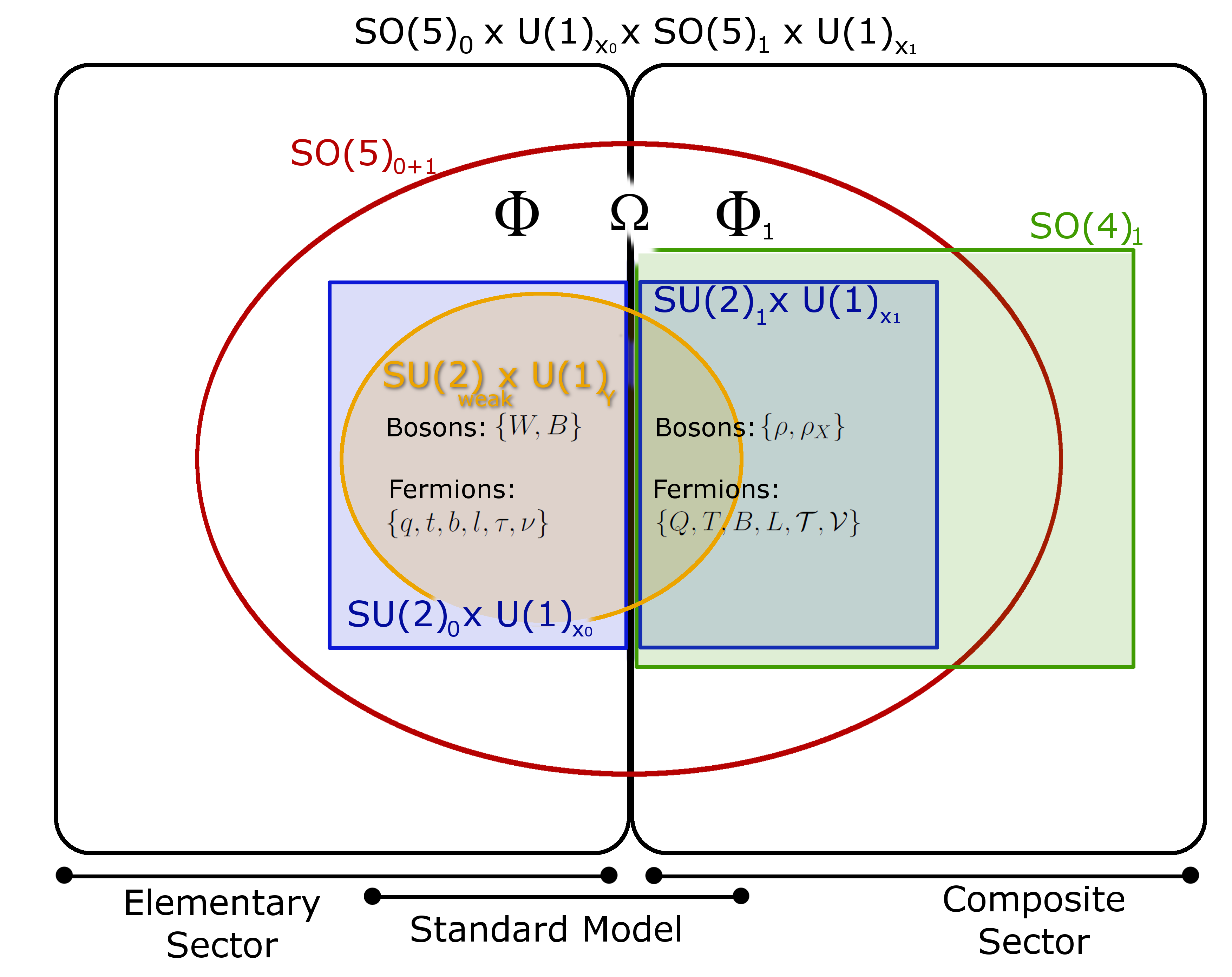}
\caption{The group structure of the two-site model considered here, with coloured-in regions denoting gauged groups. There are four sources of breaking: spontaneous $SO(5)_1 \rightarrow SO(4)_1$ by $\Phi_1$, $SO(5)_0 \times SO(5)_1 \rightarrow SO(5)_{0+1}$ by $\Omega$, explicit $SO(5)_{0+1}\rightarrow SU(2)_0 \times U(1)_{x_0}$ by removing non-dynamic fields, $SU(2)_0 \times U(1)_{x_0} \times SU(2)_1 \times U(1)_{x_1}\rightarrow SU(2)_{\textnormal{weak}} \times U(1)_Y$ by gauging the SM group}
\label{fig:twosite}
\end{figure}

The models that we consider consist of an elementary site (Site 0) and a composite site (Site 1), as summaried in Figure~\ref{fig:twosite}.  Site 1 is populated by composite partners to the elementary fields, to be thought of as the first set of resonances arising from the new strongly-coupled sector. It contains fields invariant under a global symmetry $G_1 = SO(5)_1 \times U(1)_X$, where the extra factor $U(1)_X$ turns out to be necessary to provide the correct hypercharge for the fermions. Site 0 is populated by fermion and gauge fields with the same (gauged) symmetry group and the same fermion representations as the SM, excluding the Higgs doublet. It is useful to promote the elementary symmetry group to an exact global symmetry $G'_0=SO(5)_0 \times U(1)_X$, by introducing spurious, non-dynamical gauge and fermion fields, and by temporarily assuming that all fields on the elementary site are non-dynamical. This is done purely for mathematical convenience, as it allows us to construct the Lagrangian of the low-energy effective theory by writing terms symmetric under the global product group $G_0' \times G_1$. $G_0' \times G_1$ is then spontaneously broken down to the global, diagonal subgroup $G$, and the NGBs associated with this breaking are eaten by the $G_1$ gauge fields to produce massive, vector bosons transforming in the adjoint representation of $G$. At the same time, $G_1$ spontaneously breaks to $H_1 = SO(4)_1\times U(1)_X$, producing 4 NGBs that provide a fully composite Higgs boson. The SM fields observed in nature are linear combinations of the elementary and composite source fields, i.e. they are ``partially composite".

One can now ``turn off'' the spurious fields on Site 0, and gauge only the SM $SU(2)_L\times U(1)_Y$ subgroup of $G'_0$ (leaving the SM fermions in incomplete representations of $G'_0$). The effect is to break $G$ down to the SM electroweak gauge group, but the explicit breaking is weak due to the fact that the couplings and masses on the composite site are much larger than their SM counterparts. Hence, the Higgs remains light due to its pNGB nature.



The spontaneous breaking $G_1\rightarrow H_1$ can be parameterised by a field $\Phi_1$, with a $U(1)_X$ charge of $Q_X=0$ and a non-zero vacuum expectation value of $\langle \Phi_1 \rangle=(0,0,0,0,1)$. One can write:

\begin{align}
\Phi_1(\Pi) &= \langle \Phi_1 \rangle\exp(i\Pi^{\hat{a}}T^{\hat{a}}) \label{ngb}\, ,
\end{align}
where $\Pi^{\hat{a}}T^{\hat{a}}$ contains the four NGBs for the four broken $SO(5)_1$ generators, $\{T^{\hat{a}}\}$. This field transforms as $\Phi_1 \rightarrow g_1 \Phi_1$. Site 0 and Site 1 are connected by a link field $\Omega$ that will be described by a separate $\sigma$-model, and which transforms as $\Omega \rightarrow g_0 \Omega g_1^\dagger$, $g_0, g_1 \in G'_0, G_1$. $\Omega$ parameterises the NGBs from the spontaneous breaking of $G_1\times G_0\rightarrow G$, and its addition allows the realisation of partially composite fermions through the presence in the final Lagrangian of bilinears that involve a fermion on Site 0 and a fermion on Site 1. The physical content of the theory becomes apparent when one transforms to the unitary gauge, which can be accomplished via~\footnote{For brevity, we have dropped from this discussion an extra link field $\Omega_X$ for  the $U(1)_X$ factor, but the details can be found in~\cite{carena2014}.}:
\begin{align}
\Phi &= \Omega \Phi_1\, .
\end{align}
In this gauge, the would-be NGBs that are eaten to generate the composite vector masses are set to zero. Using the $SO(5)_1$ basis $\{T^a\in H_1, T^{\hat{a}}\in G_1/H_1\}$, one can show that 

%
\begin{align}
\Phi &= \frac{1}{\hat{h}}\sin\frac{\hat{h}}{f}(h^1,h^2,h^3,h^4,\hat{h}\cot\frac{\hat{h}}{f})^T\, ,
\end{align}
where $\hat{h} = \sqrt{h^{\hat{a}}h^{\hat{a}}}$. We identify the usual Higgs doublet as
\begin{align}
H = \left(\begin{matrix}
ih_1 + h_2\\
ih_3 + h_4
\end{matrix}\right) = \left(\begin{matrix}
0\\
h
\end{matrix}\right)
\end{align}
with the usual vacuum $\langle H \rangle = (0,v)^T$. Therefore we match fields and see 
\begin{align}
h_4 = \hat{h} &= h &\implies & & \Phi &= (0,0,0,s_h,c_h)^T & \textnormal{and} & & \langle \Phi \rangle = (0,0,0,\xi,\sqrt{1-\xi^2})^T, \label{phi}
\end{align}
where $s_h = \sin\frac{h}{f}$, $c_h = \sin\frac{h}{f}$, $\xi=\sin\frac{v}{f}$ and $f$ is the NGB decay constant.

The Lagrangian for our models can be constructed by combining separate contributions that describe the elementary and composite sites, and the mixing between them. The elementary site contribution from Site 0 is given by
\begin{align}
\begin{split}
\mathcal{L}_{\rm 0} = & -\frac{1}{4g_{2,{\rm 0}}^2}W_{\mu\nu}W^{\mu\nu}-\frac{1}{4g_{1,{\rm 0}}^2}B_{\mu\nu}B^{\mu\nu}\\
& + \left(\frac{\Lambda}{d_qm_Q}\right)^2\bar{q}i\slashed{D}_{\rm 0}q+\left(\frac{\Lambda}{d_tm_T}\right)^2\bar{t}^ci\slashed{D}_{\rm 0}t^c + \left(\frac{\Lambda}{d_b m_B}\right)^2\bar{b}^ci\slashed{D}_{\rm 0}b^c\\
& + \left(\frac{\Lambda}{d_l m_L}\right)^2\bar{l}i\slashed{D}_{\rm 0}l+\left(\frac{\Lambda}{d_\tau m_\mathcal{T}}\right)^2\bar{\tau}^ci\slashed{D}_{\rm 0}\tau^c + \left(\frac{\Lambda}{d_\nu m_\mathcal{V}}\right)^2\bar{\nu}^c i\slashed{D}_{\rm 0}\nu^c\\
& + \ldots\, ,
\end{split}
\end{align}
with $\{W_{\mu\nu}, B_{\mu\nu}\}$ representing the $SU(2)_L \times U(1)_Y \in SO(5)_0 \times U(1)_0$ field strength tensors, and the matter content: $t^c$ representing the elementary right-handed top-like quark, $q$ being a left-handed doublet for the third quark generation, and so on. The covariant derivative involving the elementary gauge fields is denoted by $D_{\rm 0}$. Terms involving lighter fermions are neglected (but noted for completeness by the dots), and we also note that the quark kinetic terms do not yet follow canonical normalisation. The normalisation factor will be explained shortly. 

The composite site Lagrangian contains the terms:
\begin{align}
\begin{split}
  \label{eq:Lcomp}
\mathcal{L}_{\rm 1} = &-\frac{1}{4g_\rho^2}\rho_{\mu\nu}\rho^{\mu\nu}-\frac{1}{4g_X^2}\rho_{X,\mu\nu}\rho_X^{\mu\nu} +\frac{f_{\rm 1}^2}{2}(D_{{\rm 1},\mu}\Phi)(D_{\rm 1}^\mu\Phi)^T\\
& +\bar{Q}i\slashed{D}_{\rm 1}Q+\bar{Q}^ci\slashed{D}_{\rm 1}Q^c +\bar{T}i\slashed{D}_{\rm 1}T+\bar{T}^ci\slashed{D}_{\rm 1}T^c +\bar{B}i\slashed{D}_{\rm 1}B+\bar{B}^c i\slashed{D}_{\rm 1}B^c\\
& -m_Q \bar{Q}Q^c-m_T \bar{T}T^c -m_B \bar{B}B^c -m_{Y_T} \bar{Q}T^c -m_{Y_B} \bar{Q}B^c -Y(\Phi,Q,T,B)\\
&+ \{Q\rightarrow L, T \rightarrow \mathcal{T}, B\rightarrow \mathcal{V}\}+\mbox{h.c.}+\ldots\, ,
\end{split}
\end{align}
where $\rho$ and $\rho_X$ are the field strength tensors for the composite, $G_{\rm 1}$ gauge field; $\Phi$ contains the Higgs fields, and $D_{\rm 1}$ is the covariant derivative involving the composite gauge fields. $\{Q,T,B,L,\mathcal{T},\mathcal{V}\}$ and their charge conjugates $\{Q^c,T^c,B^c,L^c,\mathcal{T}^c,\mathcal{V}^c\}$  are the composite Dirac fermions that mix, respectively, with the elementary fields $\{q,t,b,l,\tau,\nu \}$. Three types of term can be written for the fermions: diagonal mass terms, $m_\Psi$, off-diagonal mass terms, $m_{Y_\Psi}$, and Yukawa-like terms, $Y(\Phi,Q,T,B,L,\mathcal{T},\mathcal{V})$ that couple the fermions to the Higgs. These terms are given below for each fermion representation. The need to keep the Higgs potential finite means that  $Q^cT$ terms are not present despite being allowed by all symmetries of the model.

Finally, we can write the mixing Lagrangian as:
\begin{align}
\begin{split}
\label{mixing}
\mathcal{L}_{\rm m} &= \frac{f_\Omega^2}{4}(D_{{\rm 0+1},\mu}\Omega)(D_{\rm 0+1}^\mu\Omega)^\dagger+\Lambda_q R_q(\Omega)qQ^c+ \Lambda_t R_t(\Omega)t^cT + \Lambda_b R_b(\Omega)b^c B \\
&+\Lambda_l R_l(\Omega)l L^c+ \Lambda_\tau R_\tau(\Omega)\tau^c \mathcal{T} + \Lambda_\nu R_nu(\Omega)\nu^c \mathcal{V} +\mbox{h.c.}+\ldots\, ,
\end{split}
\end{align}
where $\Omega$ is the link field defined earlier, and $D_{\rm 0+1}$ is a covariant derivative that contains both elementary and composite fields. The remaining terms mix $q$ and $Q^c$, $t^c$ and $T$, and so on, in a form that is consistent with the original $G_0\times G_1$ symmetry. This mixing is accomplished using projections, $R(\Omega)$, that correspond to the $G_1$ representations that $q, t^c, \rm etc.$ are embedded in. Since the elementary fermions are not canonically normalised, the actual couplings on the mixing terms go like $d_q m_Q$ for the $q$, $d_t m_T$ for the $t^c$ term, and so on. That is, we parameterise the elementary-composite mixing by an angle $\tan\theta_\psi = d_\psi \in [0,1]$. After we have our effective theory, it turns out to be convenient to redefine the scale of each bare mass to canonical normalisation: $\Lambda_\psi \rightarrow d_\psi m_\Psi$.

At low energies, the composite site degrees of freedom ($\rho$, $Q$, $T$, $B$, $L$, $\mathcal{T}$, $\mathcal{V}$) can be integrated out to obtain an effective theory, where momentum-dependent form factors encode the details of the composite site:
\begin{align}
\begin{split}
\label{eq:Leff}
\mathcal{L}_{\rm eff}={} & {}\frac{1}{2}P^T_{\mu\nu}\left[\Pi_W(p^2,h)W_\mu W_\nu+\Pi_B(p^2,h)B_\mu B_\nu+\Pi_{WB}(p^2,h)W^3_\mu B_\nu\right]+{}\\
& {}\Pi_t(p^2,h)\bar{t}\slashed{p}t +\Pi_b(p^2,h)\bar{b}\slashed{p}b +\Pi_{t^c}(p^2,h)\bar{t}^c\slashed{p}t^c +\Pi_{b^c}(p^2,h) \bar{b}^c \slashed{p} b^c  + \\
& M_t(p^2,h)tt^c + M_b(p^2,h)bb^c + \\ 
& \{q\rightarrow l, t\rightarrow \tau, b\rightarrow \nu\} + \mbox{h.c.} + \ldots \, ,
\end{split}
\end{align}
where $\Pi_i$ and $M$ are the form factors and $P^T$ is the transverse projection operator.  Once a choice has been made for the precise embedding of the elementary fermions, explicit expressions for the form factors can be obtained.

The one-loop Higgs potential can be shown to be
\begin{align}
\label{potential}
V(h) &= \int\limits_0^\infty \frac{\dd p^2}{16\pi^2} p^2\left( \frac{9}{2} \log\Pi_w \right) - 2 \sum\limits_{\psi=t,b,\tau,\nu} \textnormal{Nc}_\psi \int\limits_0^\infty \frac{\dd p^2}{16\pi^2}p^2 \log \left[ p^2 (1+\Pi_\psi)(1+ \Pi_{\psi^c}) - |M_\psi|^2\right]  \nonumber \\
& \equiv -\gamma s_h^2 + \beta s_h^4 \, . &
\end{align}
The second term in the first line of Eq. (\ref{potential}) is the fermion contribution to the potential, and will be discussed in the next section. It includes a factor for the number of colours of each fermion Nc$_\psi$. The potential is expanded up to quartic order in the Higgs fields, to make connection with the usual SM Higgs potential. The Higgs VEV is then given by: 
\begin{align}
\begin{split}
\xi &= \frac{\gamma}{2\beta}\\
&= \sin^2(\frac{v}{f}) \approx \frac{v^2}{f^2}
\end{split}
\end{align}
and the (composite) Higgs mass by
\begin{align}
m_h^2 = \frac{8\beta}{f^2}\xi(1-\xi)\, .
\end{align}
The mass of each SM fermion ($\psi=t,b,\tau$) can be calculated from the form factors in Eq. (\ref{potential}):
\begin{equation}
m_\psi=\frac{M_\psi(0,v)}{\sqrt{\Pi_\psi(0,v)\Pi_{\psi^c}(0,v)}}\, .
\end{equation}

In the following, we will explore three different theories that are distinguished by the choice of embedding for the leptons. For each model, we scan the composite sector parameter space to find points that reproduce measured observables. These observables are the Higgs VEV and mass, and the masses of the top quark, bottom quark and tau lepton. The tau neutrino will be treated as massless, however certain representations of the lepton composite partners can realise a see-saw model \cite{carmona2015}. In practise, the Higgs VEV only ever appears in the ratio $v^2/f^2$ and hence we can simply rescale $f$ to give the correct Higgs VEV instead of treating it as an extra input parameter. After performing this rescaling, we take the points that give correct values for the remaining observables and calculate the spectrum of predicted resonances and the expected deviations from the SM Higgs couplings. The latter are parameterised as a fraction of the composite Higgs-$\chi$-$\chi$ coupling $c$ with the SM Higgs-$\chi$-$\chi$ coupling $c_\textnormal{SM}$,
\begin{align}
r_\chi &= \frac{c(h\chi\chi)}{c_\textnormal{SM}(h\chi\chi)}\, .
\end{align}
Comparison of these predictions with current and anticipated collider results will give us limits on the fine tuning of each theory.

\subsection{Details of the gauge sector}

The gauge sector is common to each of our theories, and is unchanged from~\cite{Barnard:2015ryq,carena2014}. An angle $t_\theta\equiv \tan\theta$, assumed to be small, quantifies the amount of mixing between the elementary and composite sectors, whilst the masses of the two lightest vector resonances are given by $m_\rho$ and $m_a$. We vary these parameters in the intervals:
\begin{align}
t_\theta & \in[0,1] &
m_\rho, m_a & \in[0.5,10]\mbox{ TeV}
\end{align}
with $m_a>m_\rho$. For each point, we check that the value of $f$ found is consistent with all dimensionful parameters having magnitudes less than $4\pi f$ (both in the gauge and the fermion sectors).

Form factors in the gauge sector depend only on the symmetry breaking pattern so are the same in all models studied here. We vary $g_\rho$, $f_{\rm c}$ and $f_\Omega$ via the mixing angle and masses
\begin{align}
t_\theta & =\frac{g_{2,{\rm e}}}{g_\rho}\, , &
m_\rho^2 & =\frac{1}{2}g_\rho^2f_{\rm c}^2\, , &
m_a^2 & =\frac{1}{2}g_\rho^2(f_{\rm c}^2+f_\Omega^2)\, .
\end{align}

The form factor for the $W$ boson is

\begin{equation}
\Pi_W=-\frac{p^2(p^2-(1+t_\theta^2)m_\rho^2)}{g_2^2(1+t_\theta^2)(p^2-m_\rho^2)}+\frac{1}{4}s_h^2\left[\frac{2m_\rho^2(m_a^2-m_\rho^2)t_\theta^2}{g_2^2(1+t_\theta^2)(p^2-m_a^2)(p^2-m_\rho^2)}\right]\, ,
\end{equation}
where $g_2$ is the observed $SU(2)_L$ gauge coupling. Plugging into \eqref{potential} and performing the integral results in a contribution to the $s_h^2$ part of the Higgs potential of:
\begin{equation}
\gamma_g=-\frac{9m_\rho^4(m_a^2-m_\rho^2)t_\theta^2}{64\pi^2(m_a^2-(1+t_\theta^2)m_\rho^2)}\ln\left[\frac{m_a^2}{(1+t_\theta^2)m_\rho^2}\right]
\end{equation}
at leading order in $t_\theta$. 

The composite sector features several massive vector-boson resonances that are charged under $SU(2)_L\times U(1)_Y$. The quantum numbers and masses are given, to a very good approximation, by {\bf 1}$_{\pm 1}$ with mass $m_{\rho_1}=m_{\rho}$ and {\bf 3}$_{\pm 0}$ with mass $m_{\rho_3}=m_{\rho}/\cos\theta$. The effect is to modify the $hVV$ coupling (where $V$ is a $Z$ or $W$ boson), by
\begin{equation}
r_V=\sqrt{1-\xi}\, .
\end{equation}

There is also a correction to the loop-induced $h\gamma\gamma$ coupling, which is given by \cite{Shifman:1979eb,Carena:2012xa}:
\begin{align}
r_\gamma &= \Bigg\lvert \frac{A_1 r_V + \sum\limits_{\xi=t,b,\tau} N^c_\chi Q^2_\chi A_{1/2,\chi} r_\chi}{A_1 + \sum\limits_{\xi=t,b,\tau} N^c_\chi Q^2_\chi A_{1/2,\chi}} \Bigg\rvert
&= \Bigg\lvert \frac{A_1 r_V + \frac{4}{3} A_{1/2,t}r_t + \frac{1}{3} A_{1/2,b}r_b + A_{1/2,\tau}r_\tau}{A_1 + \frac{4}{3} A_{1/2,t} + \frac{1}{3} A_{1/2,b} + A_{1/2,\tau}}\Bigg\rvert\, ,
\end{align}
where $r_t$, $r_b$ and $r_\tau$ are the modifications to the $htt$, $hbb$ and $h\tau\tau$ couplings that we will describe in the following sections, and $A_{i,\chi}$ is the loop function for particle $\chi$ with spin $i$ and number of colours $N^c_\chi$. These are approximately~\cite{Carena:2012xa}:
\begin{align}
A_1 \approx -8.324, & & A_{1/2,t} \approx 1.375, & & A_{1/2,b} \approx -0.072 - 0.095 i, & & A_{1/2,\tau} \approx -0.024 - 0.022 i\, .
\end{align} 
The lighter fermion contributions are negligible compared to the heavier terms. They are included here for completeness, though are not included in the fine tunings given below. Finally, the modification to the $hgg$ coupling is the same as that of the $htt$ coupling if one neglects the contribution of lighter states.

\subsection{Details of the fermion sector}
\label{FermionSector}

As noted above, the specific fermion form factors that enter Eq. (\ref{eq:Leff}) depend on the way that each composite fermion is embedded in the $SO(5)_1$ group. That is, there is more than one way of representing the fermion multiplet in the Lagrangian such that it is invariant under an $SO(5)$ rotation. We are interested in all of the lowest dimension representations, the trivial ${\bf 1}$, the fundamental ${\bf 5}$, the antisymmetric ${\bf 10}$ and the symmetric traceless ${\bf 14}$. As per \cite{carena2014,carmona2015}, we will focus on three interesting ensembles of embedding: MCHM$^{\textbf{5-5-5}}_{\textbf{5-5-5}}$, MCHM$^{\textbf{5-5-5}}_{\textbf{14-14-10}}$ and MCHM$^{\textbf{5-5-5}}_{\textbf{14-1-10}}$. MCHM$^{q-t-b}_{l-\tau-\nu}$ is the $SO(5)\rightarrow SO(4)$ composite Higgs model with the lepton doublet partner in representation $l$, tau partner in representation $\tau$, and so on. We hold the quark sector fixed but vary the composite lepton representations in the definition of each model. We expect that alternate quark embeddings would give qualitatively similar results with respect to the relative differences in fine tuning for each of our lepton embedding choices (although the absolute scale may differ).

\subsubsection{MCHM$^{\textbf{5-5-5}}_{\textbf{5-5-5}}$}

We begin with the case of the new composite sector particles each embedded in the fundamental representation. In this case, we have a partner for each right-handed, third generation fermion, and two for each left-handed doublet: 
\begin{align}
Q^t, T \sim \textbf{5}_{2/3}\, , & & Q^b, B \sim \textbf{5}_{-1/3}\, , & & L^\tau, L_\nu, \mathcal{T} \sim \textbf{5}_{-1}\, , & &  \mathcal{V} \sim \textbf{5}_{0}\, ,
\label{5-5-5}
\end{align}
where $T,B,\mathcal{T},\mathcal{V}$ are the composite partners of the right-handed elementary top, bottom, tau and tau neutrino respectively. $Q^t, Q^b, L^\tau, L^\nu$ are the composite partners for the left-handed states in the third generation quark and lepton doublets of the elementary sector. It is a quirk that in $\textbf{5-5-5}$ models, we require two partners for each doublet. This is required since the decomposition of a composite fiveplet under $SO(4)\simeq SU(2)\times SU(2)$ only allows the coupling of one SM doublet (e.g. $q_L \in (\textbf{2},\textbf{2})_{2/3}$) and one SM singlet (e.g. $t_R \in \textbf{1}_{2/3}$). To couple another SM singlet (e.g. $b_R \in \textbf{1}_{-1/3}$) we must introduce a second fiveplet to preserve symmetry \cite{de2012}. We thus need to add the appropriate terms to Equation~\ref{eq:Lcomp}, i.e.  $m_{Y_T} \bar{Q}T^c \rightarrow  m_{Y_T}\bar{Q}_t T^c$, and so on. Similarly, for Equation~\ref{mixing}, we add terms such as $\bar{q} \Delta_q Q^c \rightarrow \bar{q} \Delta_{q_t} Q_t^c + \bar{q} \Delta_{q_b} Q_b^c$, and the same in the leptonic sector.

The Yukawa couplings in the composite sector are:
\begin{align}
\mathcal{L}_y &= Y_t(\bar{Q}_t \Phi)(\Phi^\dagger T^c) + Y_b (\bar{Q}_b \Phi)(\Phi^\dagger B^c) + Y_\tau (\bar{L}_\tau \Phi)(\Phi^\dagger \mathcal{T}^c) + Y_\nu(\bar{L}_\nu \Phi)(\Phi^\dagger \mathcal{V}^c)\, .
\end{align}
%
%
The remaining modifications to the Higgs couplings are now:
\begin{align}
r_\varphi = \frac{1-2\xi}{\sqrt{1-\xi}}, \varphi = {g,t,b,\tau}\, .
\end{align}

In $SU(2)_L \times U(1)_Y$ notation, the first layer of multiplets containing top-like massive resonances are

\begin{itemize}
\item ${\bf1}_{2/3}=T_{2/3}$ with mass $m_{{\bf1}_{2/3}}$;
\item ${\bf2}_{1/6}=(T_{2/3},\, B_{-1/3})$ with mass $m_{{\bf2}_{1/6}}$; and
\item ${\bf2}_{7/6}=(T_{5/3},\, B_{2/3})$ with mass $m_{{\bf2}_{7/6}}$.
\end{itemize}
More details for the model (including expressions for the form factors) are given in Appendix~\ref{app:5-5-5}.

\subsubsection{MCHM$^{\textbf{5-5-5}}_{\textbf{14-14-10}}$}
Our second case embeds the leptonic sector in the symmetric and antisymmetric representations. These give us the freedom to avoid the double-tuning present in the fundamental representation. This parametrically enlarged tuning emerges from the \textbf{5-5-5} and \textbf{10-10-10} representation structures, requiring subleading terms (i.e. $\sim \mathcal{O}(d^4_\psi)$) to provide cancellations of Higgs mass terms \cite{matsedonskyi2012, panico2012}. With a \textbf{14-14} present, we have extra quadratic invariants in the Yukawa sector, Eq. (\ref{14-14_yukawa}), that provide quadratic and quartic Higgs fields without requiring significant subleading terms for cancellation. These can be seen explicitly in the source term form factors listed in the appendix, where for \textbf{14-14-10} we have $\mathcal{O}(s^2,s^4)$ at order $\mathcal{O}(d^2_\psi)$.

We now have one partner for each SM lepton (since the lepton embedding no longer follows the 5-5-5 pattern), and the same as above for the quark sector,
\begin{align}
Q^u, T \sim \textbf{5}_{2/3}\, , & & Q^d, B \sim \textbf{5}_{-1/3}\, , & & L, \mathcal{T} \sim \textbf{14}_{-1}\, , & & \mathcal{V} \sim \textbf{10}_{0}\, .
\end{align}
The Yukawa couplings in the composite sector are
\begin{align}
\mathcal{L}_y &= Y_t(\bar{Q}_t \Phi)(\Phi^\dagger T^c) + Y_b (\bar{Q}_b \Phi)(\Phi^\dagger B^c) + Y_\tau \Phi^\dagger \bar{L} \mathcal{T}^c \Phi + \tilde{Y}_\tau (\Phi^\dagger \bar{L} \Phi)(\Phi^\dagger \mathcal{T}^c \Phi) + Y_\nu \Phi^\dagger \bar{L} \mathcal{V}^c \Phi  \label{14-14_yukawa}
\end{align}
The remaining modifications to the Higgs couplings are now
\begin{align}
r_\tau &= \frac{(6\xi - 3)y_a - 2(20\xi^2 - 23\xi +4)\tilde{y}_a}{\sqrt{1-\xi}(2(5\xi - 4)\tilde{y}_a - 3y_a)}\\
r_t, r_b, r_g &= \frac{1-2\xi}{\sqrt{1-\xi}}\, ,
\end{align}
where $y_a,\tilde{y}_a$ are the proto-Yukawa couplings for the composite tau partner, as described in the next section. Further details on this model, includuing form factor expressions, are given in Appendix~\ref{app:14-14-10}.

\subsubsection{MCHM$^{\textbf{5-5-5}}_{\textbf{14-1-10}}$}
\label{14-14-10}
Our final model embeds the lepton doublet in a \textbf{14} for the tuning reason above. However, we are now interested in seeing the effect of a fully composite tau. That is, the tau couples to a partner in the singlet representation:,
\begin{align}
Q^u, T \sim \textbf{5}_{2/3}\, , & & Q^d, B \sim \textbf{5}_{-1/3}\, , & & L \sim \textbf{14}_{-1}\, , && \mathcal{T} \sim \textbf{1}_{-1}, & &  \mathcal{V} \sim \textbf{10}_{0}\, .
\end{align}
The Yukawa couplings in the composite sector are
\begin{align}
\mathcal{L}_y &= Y_t(\bar{Q}_t \Phi)(\Phi^\dagger T^c) + Y_b (\bar{Q}_b \Phi)(\Phi^\dagger B^c) + Y_\tau (\Phi^\dagger \bar{L} \Phi)\mathcal{T}^c  + Y_\nu \Phi^\dagger \bar{L} \mathcal{V}^c \Phi\, .
\end{align}
The top partners are as above.
The modification to the Higgs couplings is now:
\begin{align}
r_\varphi = \frac{1-2\xi}{\sqrt{1-\xi}}, \varphi = {g,t,b,\tau}\, .
\end{align}
Further details can be found in Appendix~\ref{app:14-1-10}.

\section{Scan details}
\label{scanning}

The models described above have between 25 independent parameters in the MCHM$^{\textbf{5-5-5}}_{\textbf{14-1-10}}$ and 27 in the MCHM$^{\textbf{5-5-5}}_{\textbf{5-5-5}}$, that we use to derive four observables measurable at the LHC: the masses of the SM Higgs $m_H$, top quark $m_t$, bottom quark $m_b$ and tau lepton $m_\tau$. These four observables determine the likelihood of a particular parameter point.

The free parameters are:  \begin{itemize}
\item The bare masses of the lightest scalar resonances $m_\rho, m_a \in [0.5,10]$ TeV; 
\item The angle of composite-elementary mixing in the gauge sector $t_\theta\in[0,1]$; 
\item The on-diagonal bare masses of the top partners $m_{Q^u},m_{Q^d}, m_T, m_B,$ $m_L (m_{L^a}, m_{L^n},)$\footnote{\label{ftnt}These are required only for MCHM$^{\textbf{5-5-5}}_{\textbf{5-5-5}}$.} $m_\mathcal{T}, m_\mathcal{V}\in [0.5,10]$ TeV where the indices are described in the previous section;
\item The off-diagonal bare masses of the top partners $m_{y_u}, m_{y_d}, m_{y_\tau},  m_{y_n} \in [0.5,10]$ TeV; 
\item The proto-Yukawa couplings $y_t, y_b, y_\tau, (\tilde{y}_\tau)$\footnote{This is required only for MCHM$^{\textbf{5-5-5}}_{\textbf{14-10-10}}$.}$, y_\nu \in [-10,10]$ TeV; and
\item The extent to which the measured SM particles are composite $d_{q^u},d_{q^d}, d_t, d_b,$ $ d_l (d_{l^a}, d_{l^n})$\textsuperscript{\ref{ftnt}}, $d_\tau, d_\nu\in [0,1]$, where the extrema are respectively fully elementary or fully composite .
\end{itemize}

Rather than make simplifying assumptions to reduce the complexity of the parameter space (as in~\cite{carmona2015}, where a random sampling approach was used), we scan the full dimensionality of each model using the {\tt Multinest} implementation of the nested sampling technique~\cite{feroz2007, feroz2008, feroz2013}. This has proven very successful in exploring complicated, multidimensional functions encountered in a range of cosmology and particle physics examples. In order to apply it here, we first formulate the scan as a Bayesian inference problem as follows.

Given a set of input parameters, $\mathbf{x}$, we wish to obtain the region of the parameter space in which the masses of the SM fermions included in our study match the observed values. Given $\mathbf{O}\equiv \{m_h,m_t,m_b,m_\tau\}$ the likelihood of any particular model with $N_p$ parameters $\mathbf{x}$ is
\begin{equation}
p(\mathbf{O}|\mathbf{x})=\prod_a\exp\left(-\frac{[O^a(\mathbf{x})-O_{\rm exp}^a]^2}{2(\sigma^a)^2}\right)\, ,
\end{equation}
where $O^a(\mathbf{x})$ is the predicted value of the $i$th observable with experimentally measured value $O^a_{\rm exp}$, $\sigma^a$ is the error in $O^a_{\rm exp}$, and the product runs over all observables. For our purposes $\sigma^a$ characterises how close we want the masses to be to their observed values. Given a prior knowledge, $p(\mathbf{x})$, of the distribution of model parameters we can determine the posterior probability of $\mathbf{x}$ via Bayes' theorem
\begin{equation}
p(\mathbf{x}|\mathbf{O}) = \frac{p(\mathbf{O}|\mathbf{x})p(\mathbf{x})}{Z}\, .
\end{equation}
The normalisation constant, $Z$, is the Bayesian evidence
\begin{equation}\label{evidence}
Z=\int p(\mathbf{O}|\mathbf{x})(\mathbf{x}) p(\mathbf{x})d^{N_{\rm p}}\mathbf{x}\, .
\end{equation}
The nested sampling algorithm evaluates the evidence by Monte Carlo integration (after first transforming the multidimensional integral into a one-dimensional integral that can be evaluated numerically). Correctly weighted posterior samples are obtained as a by-product, and it is these samples that we use in the following sections to determine our fine tuning results. Even with the nested sampling technique, we find that the scans have very long convergence times due to a rapidly falling acceptance rate, something which is to be expected in a large volume where correctly reproducing the required observables depends on delicate cancellations between terms in complicated functions of the input parameters. The goal of our study, however, is merely to find large samples of points with the correct SM behaviour in order to analyse their behaviour. We thus do not impose strict convergence on the scans, but merely run for long enough to obtain hundreds of suitable points. We do not seek to make statistical inferences from our final results, and we use flat priors on all parameters.

The particular values for the observables $\mathcal{O}_i$ used in this scan were $\mathcal{O}_1 = m_h = 125\pm 5$GeV; $\mathcal{O}_2 = m_t = 155\pm 15$GeV; $\mathcal{O}_3 = m_b = 2.7\pm 0.5$GeV; $\mathcal{O}_4 = m_\tau = 1.8\pm 0.5$GeV; where the true observable values are assumed to be normally distributed around the predicted SM values with standard deviations as given\footnote{The values are not precisely the experimentally determined values - they have strong and electroweak RGE running applied, as outlined in \cite{xing2007}.}.

Approximately 80 million points are sampled for each model, with around 40,000 passing initial EWSB conditions. We choose to study the subset that are in the vicinity of the correct SM behaviour by applying mass cuts as follows:

\begin{align}
\{120,140,2.2,1.3\} \leq \{m_H, m_t, m_b, m_\tau\} \geq \{130,170,3.2,2.3\}\, .
\end{align}

This gives us a few hundred viable points for each model. We use each of these as the starting point for a Markov Chain Monte Carlo sampling of the same parameter space for each model, giving us a more thorough exploration of each possible preferred region. We use the Metropolis-Hastings algorithm~\cite{hastings}, with step sizes for each parameter given by 0.01 times the current value of the parameter. Our final plots use points from the Metropolis-Hastings output that pass the mass cuts.


\section{Fine-tuning in Many Observables}
\label{tuning}

Including a composite Higgs sector is a well-established method of raising the scale of natural new physics above $1$TeV. Partial compositeness of the heaviest flavour of quarks and leptons goes further to raise this scale without unsatisfying fine tuning. To consistently deal with fine-tuning comparisons between models, we would like to explore a measure that extends the usual concept of tuning to one that consistently considers every source of tuning $\Delta$. The amount of fine tuning in any particular parameter $x_i$ to produce the observable $\mathcal{O}$ has historically been that introduced by Barbieri and Giudice,
\begin{align}
\Delta_{BG,i}^\mathcal{O} = \Delta_{BG}^\mathcal{O}(x_i) = \abs*{\frac{x_i}{\mathcal{O}}\frac{\partial \mathcal{O}}{\partial x_i}}_{\mathcal{O}=\mathcal{O}_\textnormal{exp}}\label{BG}\, ,
\end{align}
where we use the definitions from the previous section. This gives a measure of fine tuning for each parameter. To find the total fine-tuning in $\mathcal{O}$, $\Delta^\mathcal{O}(\vec{x})$, one might simply take the maximum of all the $\Delta_{BG,i}$. Alternatively, one may define a vector of BG measures $\nabla^{\mathcal{O}_a}$ in the intuitive way, 
\begin{align}
\nabla^{\mathcal{O}_a} &= \left(\begin{matrix}
\Delta_{BG,1}\\
\Delta_{BG,2}\\
...\\
\Delta_{BG,n_p}
\end{matrix}\right)
\end{align}
and take the magnitude of this vector over $n_p$ parameters
\begin{align}
\Delta^\mathcal{O}(\vec{x}) = \sqrt{\sum\limits_i^{n_p} (\Delta_{BG,i})^2} = |\nabla^{\mathcal{O}_a}| \label{firstorder}\, .
\end{align}
%
%
Similarly, to extend to $n_o$ observables, $\{\mathcal{O}_a\}$, we can average over each fine-tuning
\begin{align}
\Delta_1 = \frac{1}{n_o}\sum\limits_{a=1}^{n_o}\Delta^{\mathcal{O}_a}\, .
\end{align}
This has been the state-of-the art until recently. However, as pointed out by \cite{Barnard:2015ryq}, it is often the case that these fine-tuning vectors are not aligned. That is, the fine-tuning may come from more than one source and the fine-tuning measure should reflect this special double tuning - a higher order tuning. If they are completely orthogonal, then the higher order tuning should be simply the product of each single tuning. If they are completely parallel, the higher tuning should disappear. 

For any two particular tuning vectors $\{\nabla^{\mathcal{O}_a},\nabla^{\mathcal{O}_b}\}$, a quantity displaying these criteria is 
\begin{align}
\Delta_2^{ab} = \begin{vmatrix}
\nabla^{\mathcal{O}_a}\cdot\nabla^{\mathcal{O}_a} & \nabla^{\mathcal{O}_a}\cdot \nabla^{\mathcal{O}_b}\\
\nabla^{\mathcal{O}_a}\cdot\nabla^{\mathcal{O}_b} & \nabla^{\mathcal{O}_b}\cdot \nabla^{\mathcal{O}_b}
\end{vmatrix}^\frac{1}{2}_{\mathcal{O}=\mathcal{O}_\textnormal{exp}\, .}\label{double}
\end{align}
For orthogonal tunings, $\nabla^{\mathcal{O}_a}\cdot \nabla^{\mathcal{O}_b}\rightarrow 0$ and thus $\Delta_2^{ab} \rightarrow \nabla^{\mathcal{O}_a} \nabla^{\mathcal{O}_b}$. For aligned tunings $\nabla^{\mathcal{O}_a} = \lambda \nabla^{\mathcal{O}_b}$, then $\nabla^{\mathcal{O}_a}\cdot \nabla^{\mathcal{O}_b}\rightarrow \lambda \nabla^{\mathcal{O}_a} \nabla^{\mathcal{O}_a}$ and thus $\Delta_2^{ab} \rightarrow 0$. Noting that Equation~\ref{double} is the area spanned by any two tuning vectors, this behaviour should be intuitive. 

\begin{figure}
\begin{subfigure}{0.5\textwidth}
\centering
\includegraphics[width=0.4\linewidth]{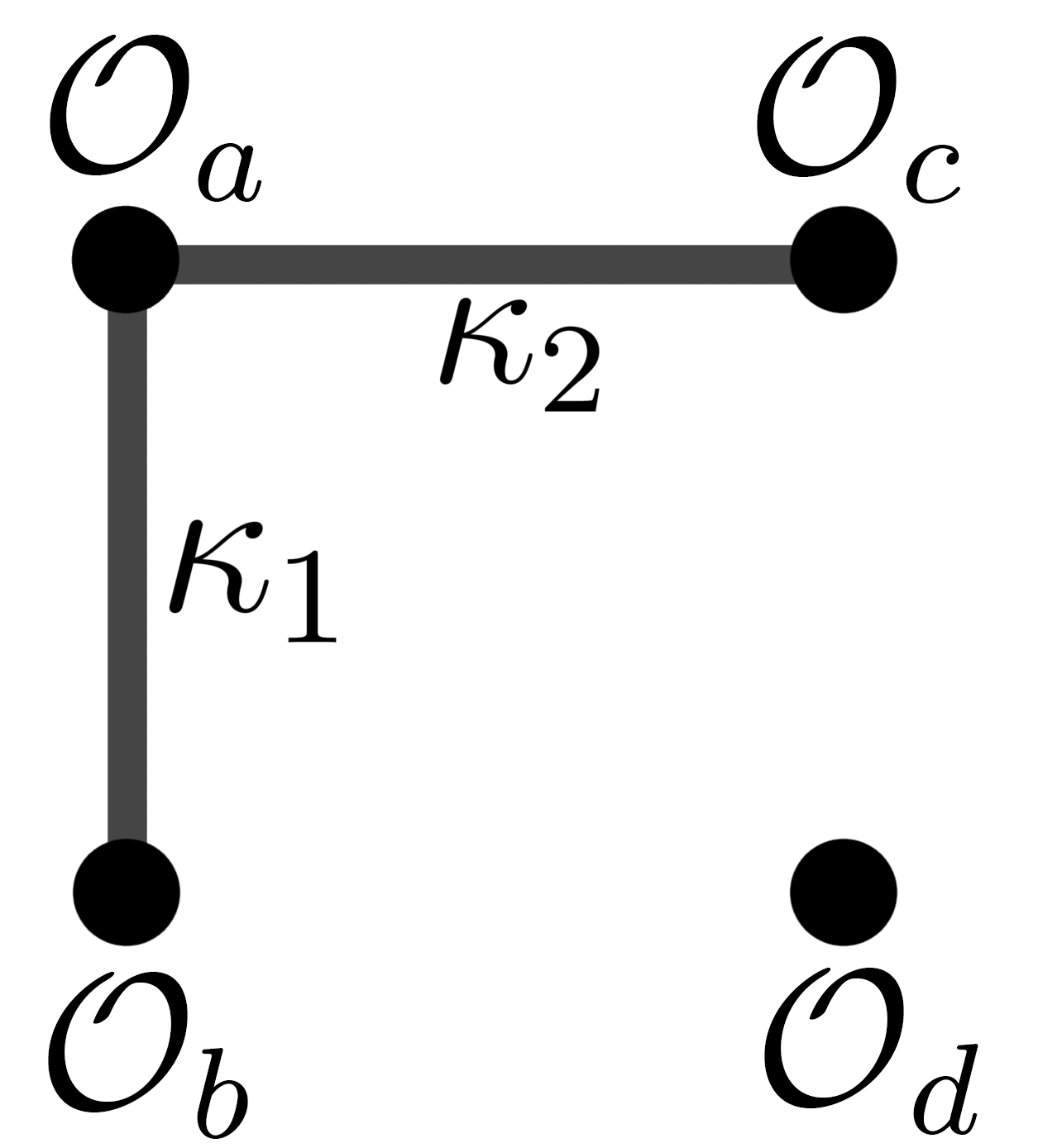}
\caption{We have one set of dependencies \\ $\mathcal{O}_a = \kappa_1\mathcal{O}_b = \kappa_2 \mathcal{O}_d$}
\end{subfigure}%
\begin{subfigure}{0.5\textwidth}
\centering
\includegraphics[width=0.4\linewidth]{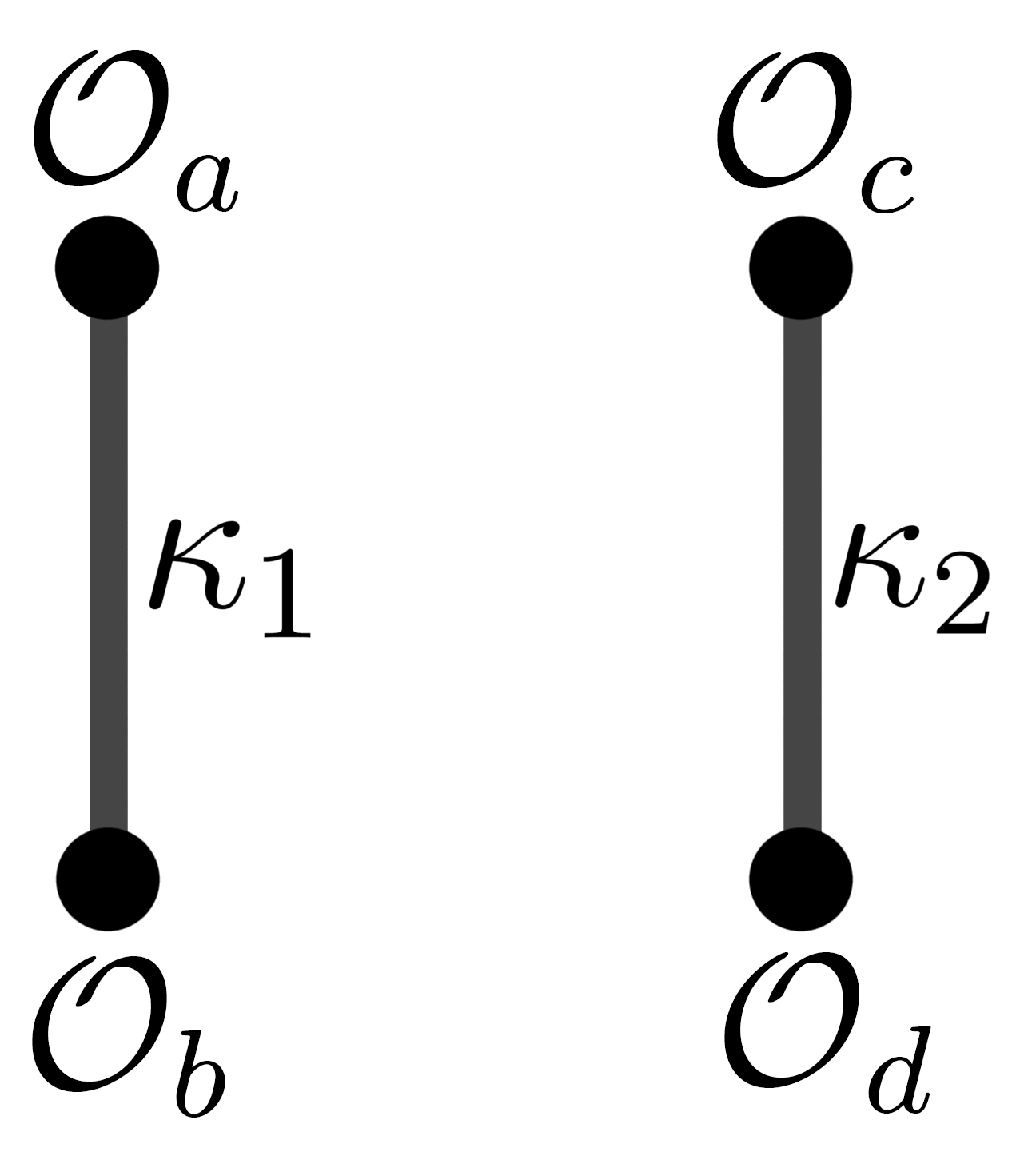}
\caption{We have two dependencies $\mathcal{O}_a = \kappa_1\mathcal{O}_b$ and $\mathcal{O}_c = \kappa_2 \mathcal{O}_d$}
\end{subfigure}
\caption{The configurations available for one source of double-tuning amongst four observables}
\label{configs}
\end{figure}

The total fine tuning $\Delta_2$ should then fulfil the criteria that (i) for all observables independent it be a maximum, (ii) for only one independent observable it vanish, and (iii) for the limiting case of two independent observables, it simply be the single double-tuning measure. For three observables, the measure satisfying these is 
\begin{align}
\Delta_2 = \frac{1}{2}(\Delta_2^{ab} +\Delta_2^{bc}+\Delta_2^{ca})\, .\label{totaldouble}
\end{align}
One can see that for observable $c$ proportional to $b$, $\mathcal{O}_c = \kappa \mathcal{O}_b$, then $\Delta_2^{bc}\rightarrow 0$ and $\Delta_2^{ac} \rightarrow \Delta_2^{ab}$. This comes from both Eq. (\ref{double}) disappearing for aligned tunings, and Eq. (\ref{BG}) being insensitive to a scaling $\kappa$.  Thus, $\Delta_2$ behaves as we would like. For more than three observables, the third criterion is not unique. In particular, for four and five observables, there are two configurations for, e.g. observables $a$ and $b$ to be independent. Configuration 1 has all dependency on one observable, configuration 2 has the dependency shared across variables, shown in figure \ref{configs}.

Configuration 1 algebraically satisfies criterion (iii) in a simple extension of Eq. (\ref{totaldouble}) to unordered pairs over $n_o$ observables
\begin{align}
\Delta_2 &= \frac{1}{n_o-1} \sum\limits_{\{\,a,b\,\} | b<a}^{\binom{n_o}{2}} \Delta_2^{ab}\label{gendoubletotal}\, .
\end{align}
However, calculating Eq. (\ref{gendoubletotal}) for configuration 2 gives more unordered pairs, and thus a factor of $4/3$ above configuration 1. This is the limit of inaccuracy in Equation~\ref{gendoubletotal}, but for randomly distributed observables the correction drops to $ <10 \% $ typically. For the purposes of fine-tuning being an order of magnitude calculation, we will accept this measure as a good approximation.

The generalisation of Equation~\ref{double} to three observables is also quite straightforward, where we take the volume spanned by three particular tuning vectors:
\begin{align}
\Delta_3^{abc} &= \begin{vmatrix}
\nabla^{\mathcal{O}_a}\cdot\nabla^{\mathcal{O}_a} & \nabla^{\mathcal{O}_a}\cdot \nabla^{\mathcal{O}_b} &\nabla^{\mathcal{O}_a}\cdot \nabla^{\mathcal{O}_c} \\\nabla^{\mathcal{O}_a}\cdot\nabla^{\mathcal{O}_b} & \nabla^{\mathcal{O}_b}\cdot \nabla^{\mathcal{O}_b} &\nabla^{\mathcal{O}_b}\cdot \nabla^{\mathcal{O}_c}\\\nabla^{\mathcal{O}_a}\cdot\nabla^{\mathcal{O}_c} & \nabla^{\mathcal{O}_b}\cdot \nabla^{\mathcal{O}_c} &\nabla^{\mathcal{O}_c}\cdot \nabla^{\mathcal{O}_c}
\end{vmatrix}^\frac{1}{2}_{\mathcal{O}=\mathcal{O}_\textnormal{exp}}\, . \label{triple}
\end{align}
Being a volume, this follows the same behaviour as the double tuning derived above. We sum various triple tunings with the extension
\begin{align}
\Delta_3 &= \frac{1}{n_o-2} \sum\limits_{\{\,a,b,c\,\} | c<b<a}^{\binom{n_o}{3}} \Delta_3^{abc}\, .
\end{align}
In general, the $N$-th order of tuning of a set of $N$ particular observables $\bm{\nabla}_N=(\nabla^{\mathcal{O}_a},\nabla^{\mathcal{O}_b},...)$ is given by 
\begin{align}
\Delta^{ab...}_N = |\bm{\nabla}_N^T \cdot\bm{\nabla}_N|^{\frac{1}{2}}
\end{align}
and the $N$-th higher order tuning over \textit{all} $n_o$ observables is 
\begin{align}
\Delta_N &= \frac{1}{n_o-(N-1)} \sum\limits_{\{\,a,b,...\,\} | ...<b<a}^{\binom{n_o}{N}} \Delta_N^{ab...}\, .
\end{align}
Finally, we simply sum each order of tuning for a measure of higher order tuning:
\begin{align}
\Delta = \sum_{i=1}^{n_o} \Delta_i\, .
\end{align}
This is the measure by which we evaluate the success of each leptonic embedding in improving the naturalness of the MCHM. Before stating and comparing our results with those of previous studies~\cite{carmona2015,Barnard:2015ryq}, it must be noted that our new measure will give larger absolute numbers for the fine tuning. To see why, consider three factors: arbitrary increase of parameters, arbitrary increase of observables, and genuinely more sensitive expressions (i.e. compare the general double tuning of Higgs mass/vev to the double tunings of the new observables). For random fine-tuning vectors, we would expect the following general dependencies. 

At order one of tuning, the number of observables $N$ will not affect the measure as they are averaged out. In terms of $n_p$ from equation \ref{firstorder}, $\Delta^\mathcal{O}$ goes as 
\begin{align}
\Delta^\mathcal{O} \sim \sqrt{n_p}\, .
\end{align}
At order two of higher order tuning - that is, double fine tuning - the measure goes as (Equations \ref{double} and \ref{totaldouble}):
\begin{align}
\Delta_2 &\propto \frac{1}{n_o -1}\left(\begin{matrix}
n_o\\
2
\end{matrix}\right) = \frac{n_o!}{(n_o - 1)2!(n_o-2)!}=\frac{n_o}{2}\sim n_o\, ,\\
\Delta_2 &\propto \begin{vmatrix}
\nabla_a \cdot \nabla_a & \nabla_a \cdot \nabla_b\\
\nabla_a \cdot \nabla_b & \nabla_b \cdot \nabla_b
\end{vmatrix}^{1/2} \sim \sqrt{\nabla_a^2 \nabla_b^2} \sim n_p\, .
\end{align}
assuming mostly orthogonal observables. That is, at second order, the measure scales linearly with number of both parameters and observables. At third order, the measure goes as 

\begin{align}
\label{tripletuning}
\Delta_3 &\propto \frac{1}{n_o -1}\left(\begin{matrix}
n_o\, ,\\
2
\end{matrix}\right) \sim n_o^2\\
\Delta_3 &\propto n_p^{3/2}\, .
\end{align}

Higher orders $\Delta_N$ follow this pattern of $\sim n_o^{N-1},n_p^{N/2}$. Of course there is a further scaling of the measure when considering higher numbers of observables. When going from three to four observables, not only do we increase the fine tuning out-of-hand by $(4/3)^2 \approx 1.8$, we also add in the possibility of order-four tuning, which is generically a factor of $\Delta_1$ greater than order-three. 

Considering all of these artefacts of the tuning measure, we arrive at a generic increase from \cite{Barnard:2015ryq} to this paper of\footnote{We remind the reader that Barnard, White have three observables and nine parameters}
\begin{align}
\textnormal{factor} &\approx	\left(\frac{5}{3}\right)^{(5-1)} \cdot \left(\frac{27}{9}\right)^{(5/2)} = 120\, .
\end{align}

\section{Results}
\label{results}

\begin{figure}
\centering
\begin{subfigure}{0.7\textwidth}
\centering
\includegraphics[width=0.9\textwidth]{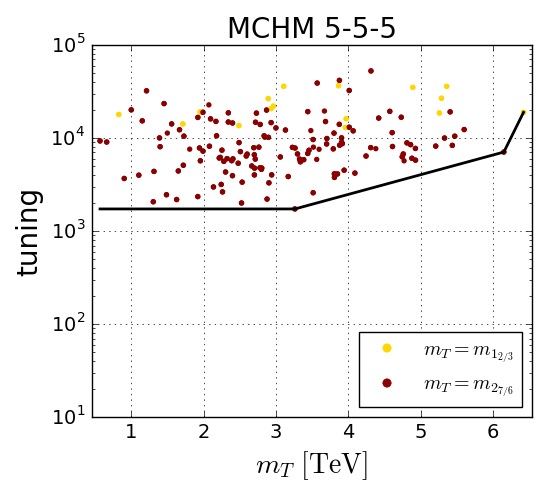}
\end{subfigure}
\begin{subfigure}{0.7\textwidth}
\centering
\includegraphics[width=0.9\textwidth]{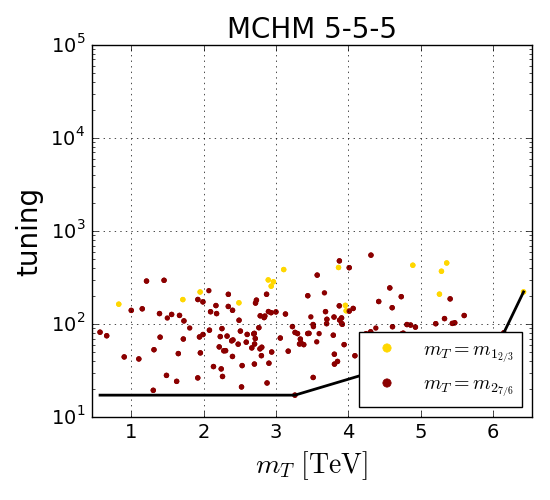}
\end{subfigure}
\caption{A comparison of non-normalised (upper) and normalised (lower) fine tunings in the mass of top partners}
\label{comparison}
\end{figure}

Below, we present the scan results in terms of the fine-tuning found at each viable parameter point. The tuning of each lepton embedding is shown against the lightest vector-boson resonance mass $m_\rho$, the lightest top partner resonance mass $m_T$, the Higgs coupling ratios $r_\chi$ and the vacuum misalignment $\xi \approx v^2/f^2$. A convex hull is provided to understand the general limits of minimal fine tuning (note that given the logarithmic scale, the hull may not always appear to be convex). We observe, in line with the prediction above, that the fine-tuning is generally two orders of magnitude higher in this lepton-sensitive case than the top-only case of \cite{Barnard:2015ryq}. If we were interested in comparing with lepton-insensitive models, for example, we could normalise by this factor. Such a normalised plot is given in Figure \ref{comparison}, along with the normalised results. For the rest of this section, we stick to using the new measure without additional normalisation, which will permit a relative comparison of our lepton embeddings (since we use the same observables in each case, and the difference between the number of parameters is not significant).

A comparison of our new tuning with less sophisticated tuning measures can be seen in the bottom right panel of Fig. (\ref{fullplots}), which shows the fine tuning for the MCHM$^{\textbf{5-5-5}}_{\textbf{5-5-5}}$ model as a function of the vacuum misalignment $\xi$. Our measure gives higher values for fine tuning relative to the single tuning $\Delta_1$ or the naive fine-tuning measure $1/\xi$, which is to be expected. In this case, with the leptons and quarks all embedded in fundamental representations of $SO(5)$, the lepton sector is not contributing much at all to the phenomenology of the model, which suffers from the double tuning effect highlighted previously. 

\subsection{MCHM$^{\textbf{5-5-5}}_{\textbf{5-5-5}}$ Fine-tuning}
\begin{figure}
\centering
\begin{subfigure}{0.45\textwidth}\label{fullrg}
\centering
\includegraphics[width=0.9\linewidth]{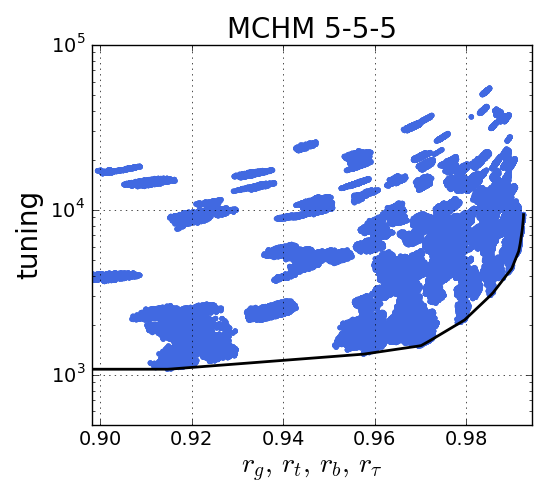}
\end{subfigure}%
\begin{subfigure}{0.45\textwidth}\label{fullrv}
\centering
\includegraphics[width=0.9\linewidth]{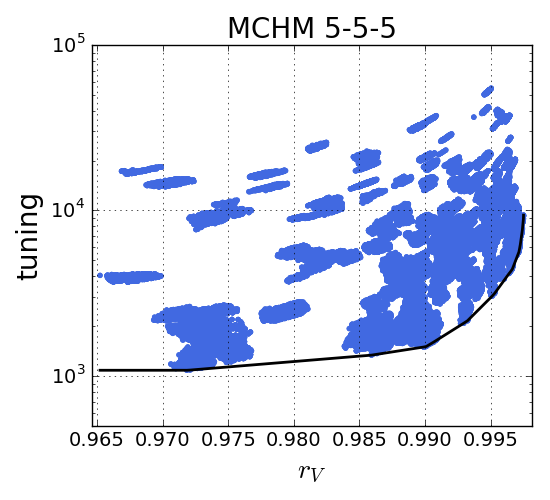}
\end{subfigure}
\begin{subfigure}{0.45\textwidth}\label{fullry}
\centering
\includegraphics[width=0.9\linewidth]{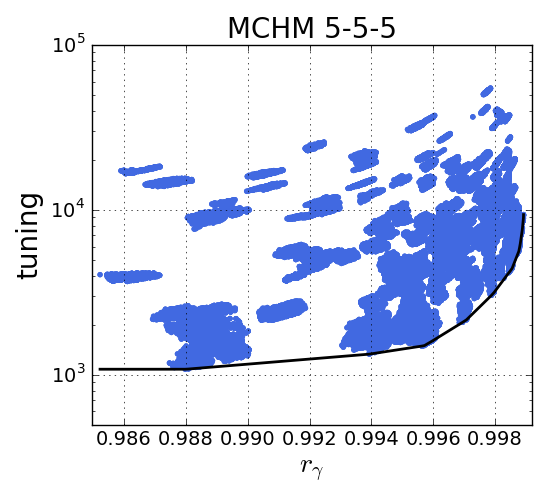}
\end{subfigure}%
\begin{subfigure}{0.45\textwidth}\label{fullmp}
\centering
\includegraphics[width=0.9\linewidth]{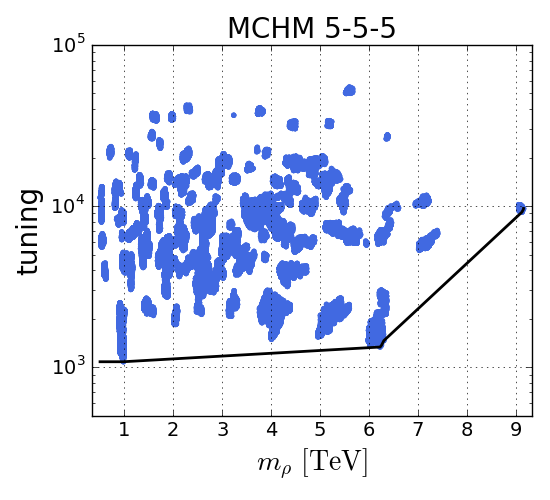}
\end{subfigure}
\begin{subfigure}{0.45\textwidth}\label{fullmt}
\centering
\includegraphics[width=0.9\linewidth]{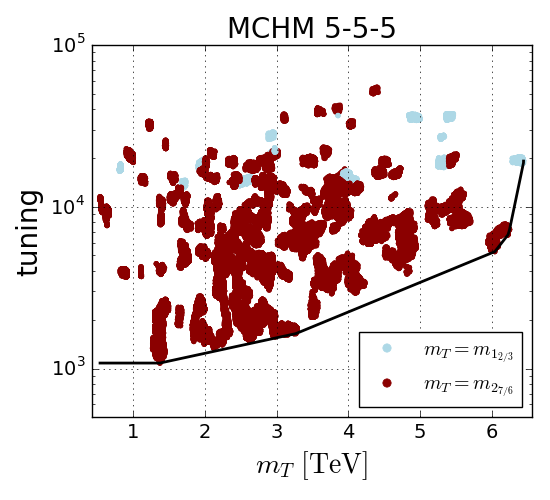}
\end{subfigure}%
\begin{subfigure}{0.45\textwidth}
\centering
\includegraphics[width=0.9\linewidth]{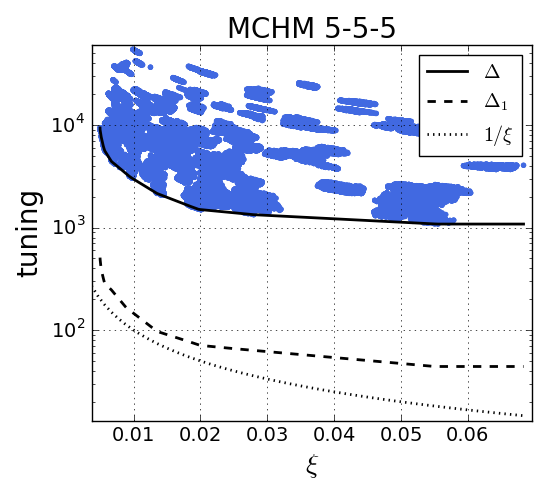}\label{fullxi}
\end{subfigure}
\caption{Tuning in the MCHM$^{\textbf{5-5-5}}_{\textbf{5-5-5}}$ model as a function of Higgs coupling ratios, lightest scalar resonance mass, top partner masses, and vacuum misalignment}
\label{fullplots}
\end{figure}

Here we present the results for the fundamental representation, found in Figure \ref{fullplots}. The full tuning is quite severe, partly due to the generic fine tuning reasons explained above, with a minimum tuning of $\Delta = 1082$ at a top partner mass of $m_{2_{7/6}} = 1.37$TeV. However, this model is particularly badly tuned, due to the quark \textit{and} lepton double tuning required to achieve EWSB. Our previous study showed a sharply linear relationship between the lightest top partner mass and the fine tuning of the point for a model that did not include the lepton sector~\cite{Barnard:2015ryq}. Our present case, however, is complicated by the fact that the inclusion of the lepton sector introduces both extra parameters and extra potential sources of tuning. These sources include the single tuning associated with reproducing the Higgs VEV and masses of the Higgs and SM fermions, and the new possibilities for multiple tunings across combinations of these observables. It is still true, however, that the fine tuning decreases with lower masses for new particles, a smaller hierarchy between elementary and composite scales, and greater divergence from Standard Model Higgs coupling predictions. There is evidence to suggest that the fine tuning rises more steeply with the lightest partner mass if this mass exceeds $3$ TeV. We also see that points for which the {\bf 2}$_{7/6}$ multiplet is the lightest top partner are significantly less finely tuned than points where it tends to be the {\bf 1}$_{2/3}$. This can be understood from the fact that the {\bf 2}$_{7/6}$ does not mix directly with the elementary top quark, and hence its mass is less constrained and easier to keep light than that of the {\bf 1}$_{2/3}$. A precision of less than 3\% on the Higgs couplings to gluons or fermions would lead to a dramatic increase in the fine tuning of the model. This precision provides the same tuning limits as excluding top partners up to $2.6$ TeV. Currently, however, Run I constraints still allow even the least finely tuned configurations. 

\subsection{MCHM$^{\textbf{5-5-5}}_{\textbf{14-14-10}}$ Fine-tuning}
\begin{figure}
\centering
\begin{subfigure}{0.45\textwidth}\label{reducedrg}
\centering
\includegraphics[width=0.9\linewidth]{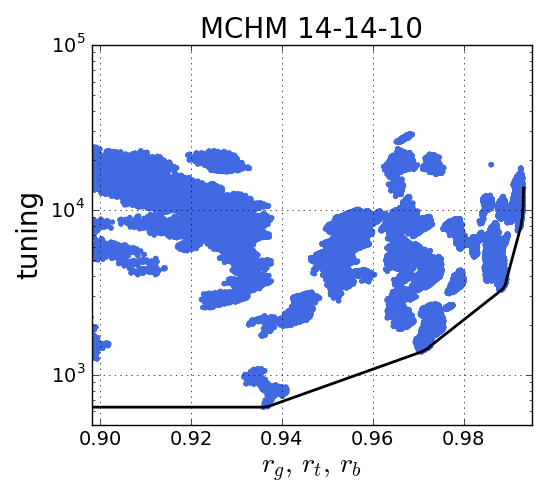}
\end{subfigure}%
\begin{subfigure}{0.45\textwidth}\label{reducedrv}
\centering
\includegraphics[width=0.9\linewidth]{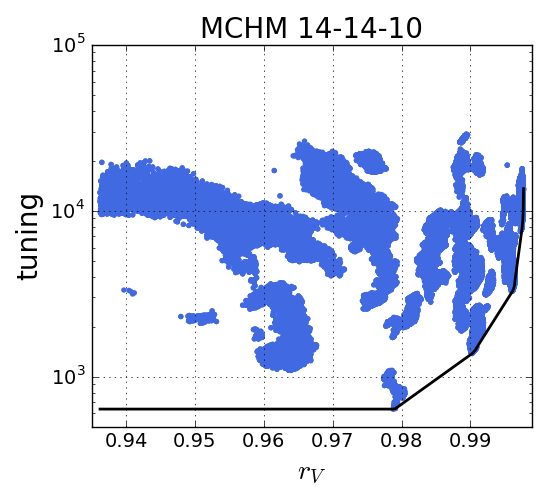}
\end{subfigure}
\begin{subfigure}{0.45\textwidth}\label{reducedry}
\centering
\includegraphics[width=0.9\linewidth]{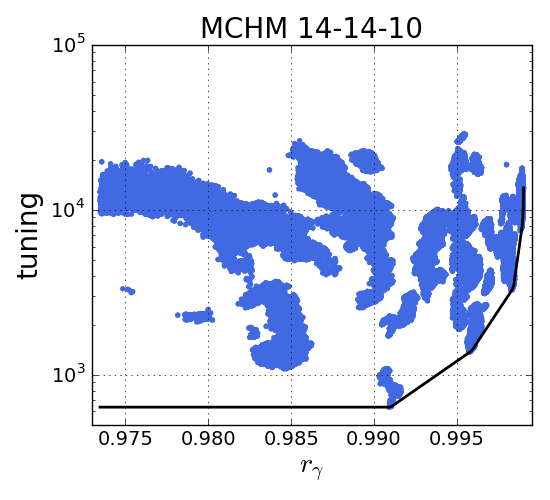}
\end{subfigure}%
\begin{subfigure}{0.45\textwidth}\label{reducedmp}
\centering
\includegraphics[width=0.9\linewidth]{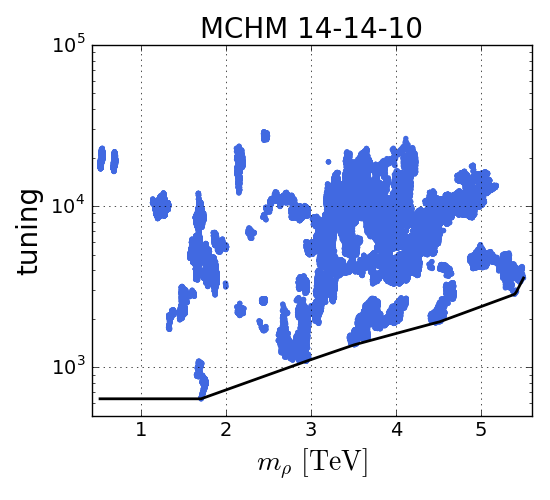}
\end{subfigure}
\begin{subfigure}{0.45\textwidth}\label{reducedmt}
\centering
\includegraphics[width=0.9\linewidth]{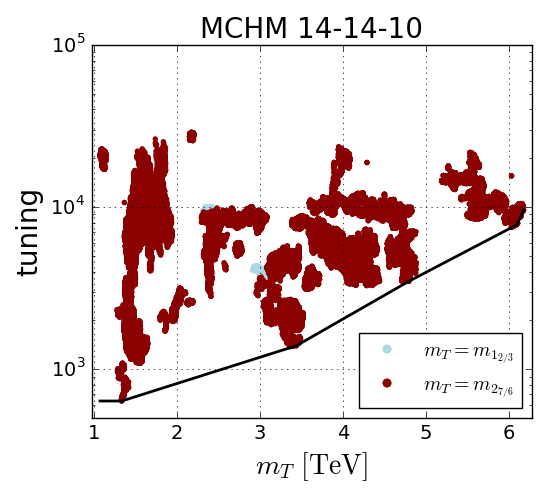}
\end{subfigure}%
\begin{subfigure}{0.45\textwidth}\label{reducedxi}
\centering
\includegraphics[width=0.9\linewidth]{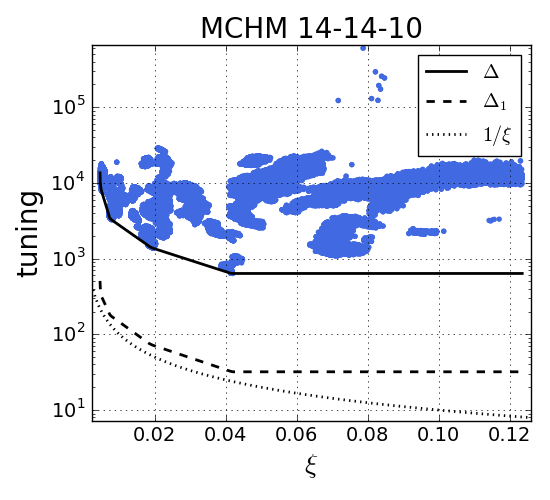}
\end{subfigure}
\caption{Tuning in the MCHM$^{\textbf{5-5-5}}_{\textbf{14-14-10}}$ model as a function of Higgs coupling ratios, lightest scalar resonance mass, top partner masses, and vacuum misalignment}
\label{reducedplots}
\end{figure}

Here we present the results for the case of symmetric representations for the leptonic doublet and the tau lepton, found in Fig. (\ref{reducedplots}). We find a much lower measure of tuning in this case than for the fundamental, which can be partly attributed to the convenient cancellation of double tuning described in Section \ref{FermionSector}. A minimum fine tuning was found to be $\Delta = 637$ at a top partner mass of $m_{2_{7/6}} = 1.34$ TeV. The fine tuning again decreases with lower masses for new particles, a smaller hierarchy in scales, and greater divergence from Standard Model Higgs coupling predictions. We see again that in the cases where the $\textbf{2}_{7/6}$ is the lightest top partner, we generally find a lower tuning.

There is evidence to suggest that, unlike in the MCHM$^{\textbf{5-5-5}}_{\textbf{5-5-5}}$ case, tuning increases more quickly for top partner masses greater than $1$ TeV. We caution, however, that the extreme difficulty of finding viable points in this model leads to a poor sampling density near the convex hull. The tuning is well below that of the MCHM$^{\textbf{5-5-5}}_{\textbf{5-5-5}}$ model for low top partner masses, but may be comparable at higher masses. Again, the reason can be attributed to the tuning measure used. Where previous works consider only the worst tuning in a particular parameter, we consider a cumulative measure that is sensitive to both the cancellation of double tuning, and the MCHM$^{\textbf{5-5-5}}_{\textbf{14-14-10}}$-specific tuning required to achieve low Higgs, top and tau masses that may be more significant at higher top partner masses. Our tuning measure also counts the increase in the number of parameters as a negative feature. Thus, although one can alleviate the double tuning in this model through organising to have a leading order contribution to the quartic Higgs potential term from the leptons, and a sub-leading contribution from the quarks, one has had to introduce additional complexity to do so, thus potentially lessening the attractiveness of the symmetric representation. A measurement of Higgs-top coupling up to 3\% would provide the same tuning constraint as excluding top partners up to $3.4$ TeV.

\begin{figure}
\centering
\includegraphics[width=0.4\linewidth]{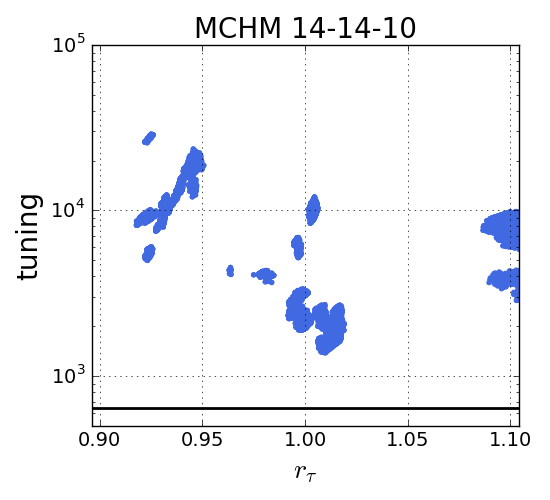}
\caption{The tuning of Higgs-tau coupling modifications}
\label{tau141410plot}
\end{figure}

Note that the Higgs-tau coupling modification has a different structure from the other models considered. In this case, the modification is much more forgiving - there exists parameter space with very little modification at low tuning. This is shown in Figure \ref{tau141410plot}.

\subsection{MCHM$^{\textbf{5-5-5}}_{\textbf{14-1-10}}$ Fine-tuning}
\begin{figure}
\centering
\begin{subfigure}{0.45\linewidth}\label{minimalrg}
\centering
\includegraphics[width=0.9\linewidth]{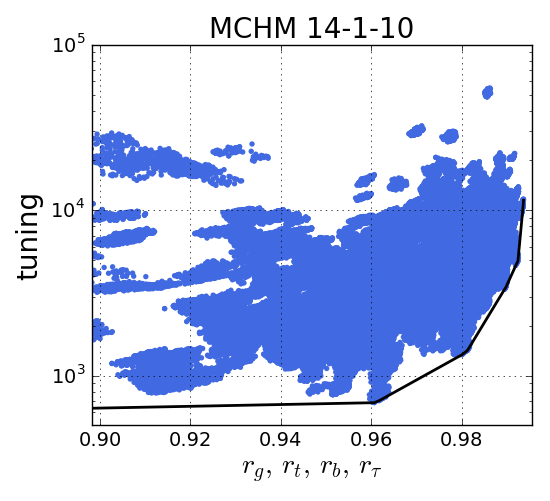}
\end{subfigure}%
\begin{subfigure}{0.45\textwidth}\label{minimalrv}
\centering
\includegraphics[width=0.9\linewidth]{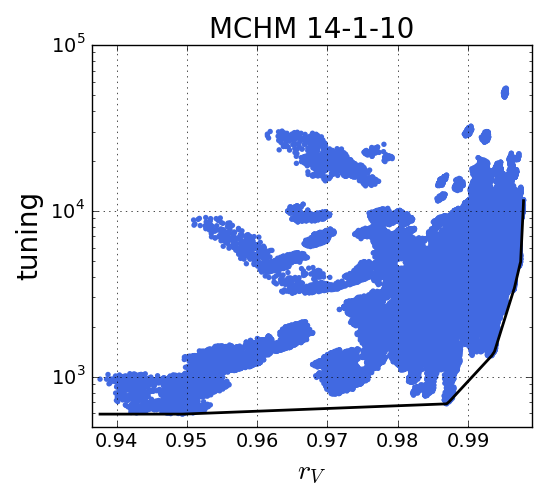}
\end{subfigure}
\begin{subfigure}{0.45\textwidth}\label{minimalry}
\centering
\includegraphics[width=0.9\linewidth]{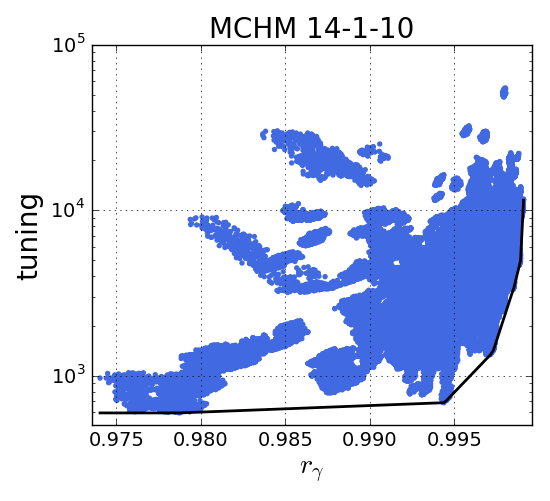}
\end{subfigure}%
\begin{subfigure}{0.45\textwidth}\label{minimalmp}
\centering
\includegraphics[width=0.9\linewidth]{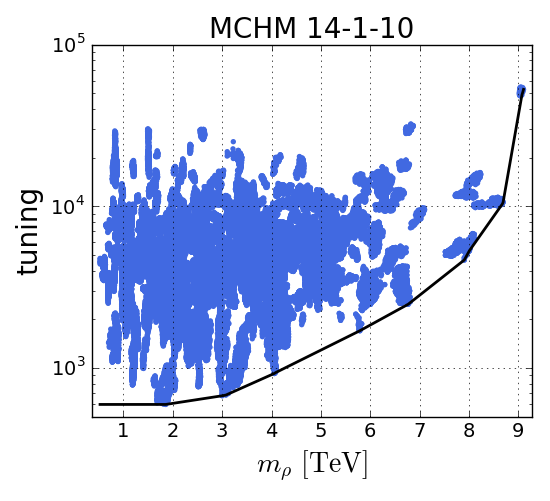}
\end{subfigure}
\begin{subfigure}{0.45\textwidth}\label{minimalmt}
\centering
\includegraphics[width=0.9\linewidth]{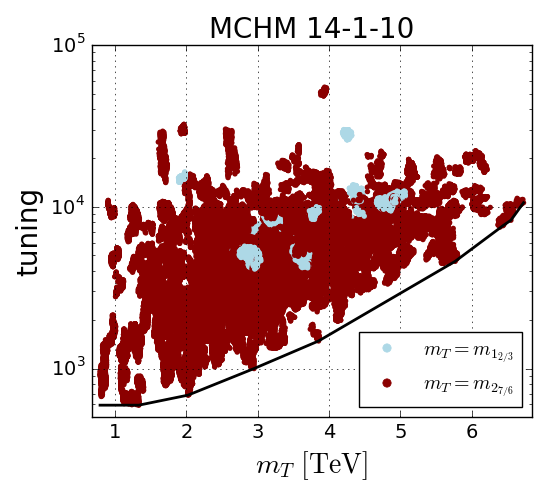}
\end{subfigure}%
\begin{subfigure}{0.45\textwidth}\label{minimalxi}
\centering
\includegraphics[width=0.9\linewidth]{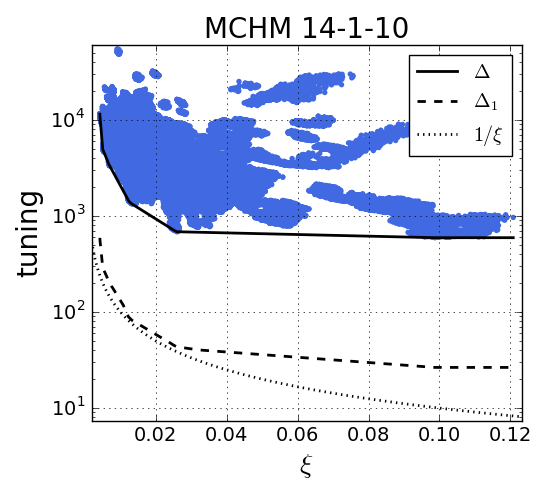}
\end{subfigure}
\caption{Tuning in the MCHM$^{\textbf{5-5-5}}_{\textbf{14-1-10}}$ model as a function of Higgs coupling ratios, lightest scalar resonance mass, top partner masses, and vacuum misalignment}
\label{minimalplots}
\end{figure}

Finally, we show the results for the case of a fully composite tau lepton, found in Figure \ref{minimalplots}. The tuning is similar to the previous case, with a minimum tuning of $\Delta = 594$ at a top partner mass of $m_{2_{7/6}} = 1.37$ TeV. However, where a natural symmetric representation shows a sharp rise in the fine tuning with better top partner mass exclusion limits and more precise Higgs coupling measurements, the present model remains relatively untuned even at top partner masses of $m_{2_{7/6}} = 3.3$ TeV, which corresponds to a coupling ratio precision of $r_\psi \approx 2\%$ (in the MCHM$^{\textbf{5-5-5}}_{\textbf{14-1-10}}$, Higgs coupling modifications have identical fine-tunings regardless of the species of particle being coupled with). This leaves the fully composite tau scenario as the likely most-natural representation once further Run II data is released. 

\section{Conclusions}
\label{sec:conclusions}
We have performed comprehensive scans to study the fine tuning of three Minimal Composite Higgs scenarios with realistic lepton sectors, distinguished by the choice of lepton embeddings. In doing so, we have had to develop a new fine tuning measure that counts the expanded range of single and multiple tunings that can occur in these scenarios. We find that the resulting measure scales with the number of observables and parameters in any given problem, and thus naturally penalises additional model complexity.

To deal with the significantly large parameter spaces encountered in MCHM models with leptons included, we developed a sophisticated sampling approach based on a combination of nested sampling and Markov Chain Monte Carlo sampling. Even with this, it proves extremely difficult to find viable points for study in these models, typically requiring weeks of cluster running in each case. The third generation quark doublet, right-handed top and right-handed bottom are all assumed to be embedded in the fundamental representation of $SO(5)$. In line with previous work, we find that there is some advantage to embedding at least one of the leptons in a {\bf 14} representation of $SO(5)$, although the effect is less dramatic than previous studies due to the complexity cost built in to the new fine tuning measure. Embedding each of the third generation lepton doublet, right-handed tau and right-handed tau neutrino in fundamental representations of $SO(5)$ leads to a minimum fine tuning of $\Delta = 1082$, which is expected to increase with top partner mass limits at the LHC, and better collider measurements of the Higgs couplings. This can be compared with a minimum tuning of $\Delta = 637$ for the MCHM$^{\textbf{5-5-5}}_{\textbf{14-14-10}}$ and a minimum tuning of $\Delta = 594$ for the MCHM$^{\textbf{5-5-5}}_{\textbf{14-1-10}}$ model. The absolute value of these fine tunings is significantly worse than previous quoted values due to the new measure, but one may choose to normalise out the complexity cost at each order of tuning, as given in the equation following Eq. (\ref{tripletuning}). In this case, the current minimum fine tunings for the MCHM$^{\textbf{5-5-5}}_{\textbf{5-5-5}}$, MCHM$^{\textbf{5-5-5}}_{\textbf{14-14-10}}$ and MCHM$^{\textbf{5-5-5}}_{\textbf{14-1-10}}$ scenarios are approximately $10\%$, $20\%$ and $20\%$  respectively. Note that these optimistic tunings should be compared to tunings of other models that have also been appropriately normalised.

Finally, it is interesting to note that our explored models behave differently with respect to future improvements in collider measurements. The MCHM$^{\textbf{5-5-5}}_{\textbf{14-14-10}}$ scenario, although currently less fine-tuned than the MCHM$^{\textbf{5-5-5}}_{\textbf{5-5-5}}$, will look similarly unnatural once Higgs coupling measurements of the fermion decay channels reach a precision of 3\%, or top partner exclusion limits reach a mass of $3.4$ TeV. The MCHM$^{\textbf{5-5-5}}_{\textbf{14-1-10}}$ scenario, meanwhile, enjoys a relatively low increase in fine tuning, even up to Higgs coupling limits of 2\%, or top partner limits of $3.3$ TeV. Higgs coupling limits and top partner limits provide complementary probes for the naturalness of composite Higgs scenarios in the next decade.


\section*{Acknowledgements}

We thank Peter Stangl for helpful comments. The work of M.J.W is supported by the Australian Research Council Future Fellowship FT140100244. D.T.M A.G.W. are supported by the ARC Centre of Excellence for Particle Physics at the Terascale (CoEPP) (CE110001104) and the Centre for the Subatomic Structure of Matter (CSSM). D.T.M is supported by an Australian Government Research Training Program (RTP) Scholarship.


\appendix

\section{Fermion Representation Expressions}

The source term form factors implicitly defined in Equations \ref{eq:Leff} and \ref{potential} can be written in terms of the decomposed form factor expressions \ref{app:5-5-5}, \ref{app:14-14-10}, \ref{app:14-1-10}. Each representation's form factors generally depend on the four following functions:
\begin{align}
\begin{split}\label{formulas}
A_L(m_1,m_2,m_3,m_4,\Delta) &= \Delta^2\left(m_1^2 m_2^2 + m_1^2 m_4^2 + m_2^2 m_3^2 - p^2(m_1^2 + m_2^2 + m_3^2 +m_4^2) + p^4\right) \\
A_R(m_1,m_2,m_3,m_4,\Delta) &= \Delta^2 \left( m_1^2 m_2^2 + m_2^2 m_3^2 - p^2(m_1^2 + m_2^2 + m_3^2 + m_4^2) + p^4\right)\\
A_M(m_1,m_2,m_3,m_4,\Delta_1,\Delta_2) &= \Delta_1 \Delta_2 m_1 m_2 m_4 (m_3^2 - p^2)\\
B(m_1,m_2,m_3,m_4,m_5) &= m_1^2 m_2^2 m_3^2 - p^2 \left(m_1^2 m_2^2 + m_1^2 m_3^2 + m_2^2 m_3^2 + m_2^2 m_5^2 + m_3^2 m_4^2\right) \\
&+ p^4\left( m_1^2 + m_2^2 + m_3^2 + m_4^2 + m_5^2\right) - p^6
\end{split}
\end{align} 
The precise expressions for the source terms in this study are slightly different from both \cite{Barnard:2015ryq,carena2014}, so we present them in full for each representation. The expressions for the $SO(4)$ decomposed form factors are to be found originally in \cite{carena2014}. They are included here for completeness.

\subsection{MCHM$^{\textbf{5-5-5}}_{\textbf{5-5-5}}$}
\label{app:5-5-5}

\textbf{Top quark:}
\begin{align*}
\Pi_{t} &= \frac{\Delta_{q_t}^2}{(m_{Q_t} d_{Q_t})^2} + \hat{\Pi}_{q_t}^{(4)} + \hat{\Pi}_{q_b}^{(4)} + \frac{s_h^2}{2}\left( \hat{\Pi}_{q_t}^{(1)} -\hat{\Pi}_{q_t}^{(4)}\right)\\
\Pi_{t^c} &= \frac{\Delta_t^2}{(m_T d_T)^2} + \hat{\Pi}_t^{(4)} + (1-s_h^2)\left( \hat{\Pi}_t^{(1)} - \hat{\Pi}_t^{(4)}\right)\\
M_t &= \frac{1}{\sqrt{2}}s_h \sqrt{1-s_h^2}\left( \hat{M}_t^{(1)} - \hat{M}_t^{(4)}\right)
\end{align*}
\textbf{Bottom quark:}
\begin{align*}
\Pi_{b} &= \frac{\Delta_{q_b}^2}{(m_{Q_b} d_{Q_b})^2} + \hat{\Pi}_{q_t}^{(4)} + \hat{\Pi}_{q_b}^{(4)} + \frac{s_h^2}{2}\left( \hat{\Pi}_{q_b}^{(1)} -\hat{\Pi}_{q_b}^{(4)}\right)\\
\Pi_{b^c} &= \frac{\Delta_b^2}{(m_B d_B)^2} + \hat{\Pi}_b^{(4)} + (1-s_h^2)\left( \hat{\Pi}_b^{(1)} - \hat{\Pi}_b^{(4)}\right)\\
M_b &= \frac{1}{\sqrt{2}}s_h \sqrt{1-s_h^2}\left( \hat{M}_b^{(1)} - \hat{M}_b^{(4)}\right)
\end{align*}
\textbf{Tau lepton:}
\begin{align*}
\Pi_{\tau} &= \frac{\Delta_{l_\tau}^2}{(m_{L_\tau} d_{L_\tau})^2} + \hat{\Pi}_{l_\tau}^{(4)} + \hat{\Pi}_{l_\nu}^{(4)} + \frac{s_h^2}{2}\left( \hat{\Pi}_{l_\tau}^{(1)} -\hat{\Pi}_{l_\tau}^{(4)}\right)\\
\Pi_{\tau^c} &= \frac{\Delta_\tau^2}{(m_\mathcal{T} d_\mathcal{T})^2} + \hat{\Pi}_\tau^{(4)} + (1-s_h^2)\left( \hat{\Pi}_\tau^{(1)} - \hat{\Pi}_\tau^{(4)}\right)\\
M_\tau &= \frac{1}{\sqrt{2}}s_h \sqrt{1-s_h^2}\left( \hat{M}_\tau^{(1)} - \hat{M}_\tau^{(4)}\right)
\end{align*}
\textbf{Tau neutrino lepton:}
\begin{align*}
\Pi_{\nu} &= \frac{\Delta_{l_\nu}^2}{(m_{L_\nu} d_{L_\nu})^2} + \hat{\Pi}_{l_\nu}^{(4)} + \hat{\Pi}_{l_\tau}^{(4)} + \frac{s_h^2}{2}\left( \hat{\Pi}_{l_\nu}^{(1)} -\hat{\Pi}_{l_\nu}^{(4)}\right)\\
\Pi_{\nu^c} &= \frac{\Delta_\nu^2}{(m_\mathcal{N} d_\mathcal{N})^2} + \hat{\Pi}_\nu^{(4)} + (1-s_h^2)\left( \hat{\Pi}_\nu^{(1)} - \hat{\Pi}_\nu^{(4)}\right)\\
M_\nu &= \frac{1}{\sqrt{2}}s_h \sqrt{1-s_h^2}\left( \hat{M}_\nu^{(1)} - \hat{M}_\nu^{(4)}\right)
\end{align*}
with the $SO(4)$ decomposed form factors given by
\begin{equation}
\begin{aligned}\label{eq:broken5-5-5}
\hat{\Pi}_{q_{t/b}}^{(1)} &= \frac{A_L(m_{T/B}, 0, m_{Y_{T/B}} + Y_{T/B}, 0, \Lambda_{q_{t/b}})}{B(m_{Q_{t/b}}, m_{T/B}, 0, m_{Y_{T/B}} + Y_{T/B}, 0)}, & \hat{\Pi}_{q_{t/b}}^{(4)} &= \frac{A_L(m_{T/B}, 0, m_{Y_{T/B}}, 0, \Lambda_{q_{t/b}})}{B(m_{Q_{t/b}}, m_{T/B}, 0, m_{Y_{T/B}}, 0)}\\
\hat{\Pi}_{t/b}^{(1)} &= \frac{A_R(m_{Q_{t/b}}, 0, m_{Y_{T/B}} + Y_{T/B}, 0, \Lambda_{t/b})}{B(m_{Q_{t/b}}, m_{T/B}, 0, m_{Y_{T/B}} + Y_{T/B}, 0)}, & \hat{\Pi}_{t/b}^{(4)} &= \frac{A_R(m_{Q_{t/b}}, 0, m_{Y_{T/B}}, 0, \Lambda_{t/b})}{B(m_{Q_{t/b}}, m_{T/B}, 0, m_{Y_{T/B}}, 0)}\\
\hat{M}_{t/b}^{(1)} &= \frac{A_M(m_{Q_{t/b}}, m_{T/B}, 0, m_{Y_{T/B}} + Y_{T/B}, \Lambda_{q_{t/b}}, \Lambda_{t/b})}{B(m_{Q_{t/b}}, m_{T/B}, 0, m_{Y_{T/B}} + Y_{T/B}, 0)}, \\
\hat{M}_{t/b}^{(4)} &= \frac{A_M(m_{Q_{t/b}}, m_{T/B}, 0, m_{Y_{T/B}}, \Lambda_{q_{t,b}}, \Lambda_{t/b})}{B(m_{Q_{t/b}}, m_{T/B}, 0, m_{Y_{T/B}}, 0)}
\end{aligned}
\end{equation}

The same expressions apply for the leptonic form factors, with the substitutions $q \rightarrow l, t\rightarrow \tau, b\rightarrow \nu$.

\subsection{MCHM$^{\textbf{5-5-5}}_{\textbf{14-14-10}}$}
\label{app:14-14-10}

The quark expressions are as above.

\textbf{Tau lepton:}
\begin{align*}
\Pi_{\tau} &= \frac{\Delta_l^2}{(m_L d_L)^2} + \hat{\Pi}_l^{(9)} + \left(\hat{\Pi}_l^{(4)} -\hat{\Pi}_l^{(4)}\right)\left(1-\frac{s_h^2}{2}\right) + \frac{1}{4}s_h^2(1-s_h^2)\left(5\hat{\Pi}_l^{(1)} - 8 \hat{\Pi}_l^{(4)} + 4\hat{\Pi}_l^{(9)}\right)\\
\Pi_{\tau^c} &= \frac{\Delta_\tau}{(m_\mathcal{T} d_\mathcal{T})^2} + \hat{\Pi}_\tau^{(9)} + 2\left(\hat{\Pi}_\tau^{(4)} -\hat{\Pi}_\tau^{(9)}\right) \left( \frac{4}{5} - \frac{3}{4}s_h^2 \right) + \frac{1}{5}(4 - 5s_h^2)^2 \left( 5\hat{\Pi}_\tau^{(1)} - 8\hat{\Pi}_\tau^{(4)} + 3\hat{\Pi}_\tau^{(9)}\right) \\
M_\tau &= \frac{3 i}{2\sqrt{5}}\left( \hat{M}_\tau^{(4)} - \hat{M}_\tau^{(9)}\right) s_h \sqrt{1-s_h^2} + \frac{i}{8\sqrt{5}}(4-5 s_h^2)\left(5\hat{M}_\tau^{(1)} - 8\hat{M}_\tau^{(4)} + 3\hat{M}_\tau^{(9)}\right)
\end{align*}
\textbf{Tau neutrino lepton:}
\begin{align*}
\Pi_{\nu} &= \frac{\Delta_l^2}{(m_L d_L)^2} + \hat{\Pi}_l^{(9)} + (1-s_h^2)\left(\hat{\Pi}_l^{(4)} - \hat{\Pi}_l^{(9)}\right)\\
\Pi_{\nu^c} &= \frac{\Delta_\nu^2}{(m_\mathcal{N} d_\mathcal{N})^2} + \hat{\Pi}_\nu^{(6)} + \frac{1}{2}s_h^2 \left( \hat{\Pi}_\nu^{(4)} - \hat{\Pi}_\nu^{(6)}\right)\\
M_\nu &= \frac{-1}{\sqrt{2}}s_h \sqrt{1-s_h^2}\hat{M}_\nu^{(4)}
\end{align*}
with the $SO(4)$ decomposed form factors given by
\begin{equation}
\begin{aligned}\label{eq:broken14-14-10}
\hat{\Pi}_l^{(9)} &= \frac{A_L(m_\mathcal{T},0,m_{Y_\mathcal{T}},0,\Lambda_l)}{B(m_L,m_\mathcal{T},0,m_{Y_\mathcal{T}},0)}, & \hat{\Pi}_l^{(4)} &= \frac{A_L(m_\mathcal{T},m_\mathcal{V},m_{Y_\mathcal{T}} + Y_\mathcal{T}/2, Y_\mathcal{V}/2, \Lambda_l)}{B(m_L, m_\mathcal{T},m_\mathcal{V},m_{Y_\mathcal{T}} + Y_\mathcal{T}/2, Y_\mathcal{V}/2)},\\
\hat{\Pi}_l^{(1)} &= \frac{A_L(m_\mathcal{T}, 0, m_{Y_\mathcal{T}} + (Y_\mathcal{T}+\tilde{Y}_\mathcal{T})4/5,0,\Lambda_l)}{B(m_L,m_\mathcal{T},0,m_{Y_\mathcal{T}} + (Y_\mathcal{T} + \tilde{Y}_\mathcal{T})4/5,0)}\\
\hat{\Pi}_\tau^{(9)} &= \frac{A_R(m_L, 0, m_{Y_\mathcal{T}},0,\Lambda_\tau)}{B(m_L, m_\mathcal{T},0,m_{Y_\mathcal{T}},0)}, & 
\hat{\Pi}_\tau^{(4)} &= \frac{A_R(m_L,m_\mathcal{V},m_{Y_\mathcal{T}} + Y_\mathcal{T}/2, Y_\mathcal{V}/2, \Lambda_\tau)}{B(m_L, m_\mathcal{T},m_\mathcal{V},m_{Y_\mathcal{T}} + Y_\mathcal{T}/2, Y_\mathcal{V}/2)},\\
\hat{\Pi}_\tau^{(1)} &= \frac{A_R(m_L, 0, m_{Y_\mathcal{T}} + (Y_\mathcal{T}+\tilde{Y}_\mathcal{T})4/5,0,\Lambda_\tau)}{B(m_L,m_\mathcal{T},0,m_{Y_\mathcal{T}} + (Y_\mathcal{T} + \tilde{Y}_\mathcal{T})4/5,0)}\\
\hat{\Pi}_\nu^{(4)} &= \frac{A_R(m_L, m_\mathcal{T},Y_\mathcal{V}/2,m_{Y_\mathcal{T}}+Y_\mathcal{T}/2,\Lambda_\nu)}{B(m_L,m_\mathcal{T},m_\mathcal{V},m_{Y_\mathcal{T}} + Y_\mathcal{T}/2,Y_\mathcal{V}/2)}, & \hat{\Pi}_\nu^{(6)} &= \frac{A_R(m_L,0,0,0,\Lambda_\nu)}{B(m_L,m_\mathcal{V},0,0,0)},\\
\hat{M}_\tau^{(9)}&= \frac{A_M(m_L, m_\mathcal{T},0,m_{Y_\mathcal{T}}, \Lambda_l, \Lambda_\tau)}{B(m_L,m_\mathcal{T},0,m_{Y_\mathcal{T}},0)}, & \hat{M}_\tau^{(4)} &= \frac{A_M(m_L,m_\mathcal{T},m_\mathcal{V},m_{Y_\mathcal{T}} + Y_\mathcal{T}/2,\Lambda_l, \Lambda_\tau)}{B(m_L,m_\mathcal{T},m_\mathcal{V},m_{Y_\mathcal{T}} + Y_\mathcal{T}/2, Y_\mathcal{V}/2)},\\
\hat{M}_\tau^{(1)} &= \frac{A_M(m_L, m_\mathcal{T}, 0, m_{Y_\mathcal{T}} + (Y_\mathcal{T} + \tilde{Y}_\mathcal{T})4/5, \Lambda_l, \Lambda_\tau)}{B(m_L, m_\mathcal{T}, 0, m_{Y_\mathcal{T}} + (Y_\mathcal{T} + \tilde{Y}_\mathcal{T})4/5, 0)},\\
\hat{M}_\nu^{(4)} &= -i \frac{A_M(m_L, m_\mathcal{V}, m_\mathcal{T}, Y_\mathcal{V}/2, \Lambda_l, \Lambda_\nu)}{B(m_L, m_\mathcal{T}, m_\mathcal{V}, m_{Y_\mathcal{T}} +Y_\mathcal{T}/2, Y_\mathcal{V}/2)}
\end{aligned}
\end{equation}

\subsection{MCHM$^{\textbf{5-5-5}}_{\textbf{14-1-10}}$}
\label{app:14-1-10}

The quark expressions are as above.

\textbf{Tau lepton:}
\begin{align*}
\Pi_{\tau} &= \frac{\Delta_l^2}{(m_L d_L)^2} + \hat{\Pi}_l^{(9)} + \left(\hat{\Pi}_l^{(4)} -\hat{\Pi}_l^{(4)}\right)\left(1-\frac{s_h^2}{2}\right) + \frac{1}{4}s_h^2(1-s_h^2)\left(5\hat{\Pi}_l^{(1)} - 8 \hat{\Pi}_l^{(4)} + 4\hat{\Pi}_l^{(9)}\right)\\
\Pi_{\tau^c} &= \frac{\Delta_\nu}{(m_\mathcal{T} d_\mathcal{T})^2} + \hat{\Pi}_\tau^{(1)}\\
M_\tau &= \frac{-\sqrt{5}}{4} s_h \hat{M}_\tau^{(1)}
\end{align*}
\textbf{Tau neutrino lepton:}
\begin{align*}
\Pi_{\nu} &= \frac{\Delta_l^2}{(m_L d_L)^2} + \hat{\Pi}_l^{(9)} + (1-s_h^2)\left(\hat{\Pi}_l^{(4)} - \hat{\Pi}_l^{(9)}\right)\\
\Pi_{\nu^c} &= \frac{\Delta_\nu^2}{(m_\mathcal{V} d_\mathcal{V})^2} + \hat{\Pi}_\nu^{(6)} + \frac{1}{2}s_h^2 \left( \hat{\Pi}_\nu^{(4)} - \hat{\Pi}_\nu^{(6)}\right)\\
M_\nu &= \frac{-i}{\sqrt{2}}s_h \sqrt{1-s_h^2}\hat{M}_\nu^{(4)}
\end{align*}
with the $SO(4)$ decomposed form factors given by
\begin{equation}
\begin{aligned}\label{eq:broken14-1-10}
\hat{\Pi}_l^{(9)} &= \frac{A_L(0,0,0,0,\Lambda_l)}{B(m_L,0,0,0,0)}, & \hat{\Pi}_l^{(4)} &= \frac{A_L(0,m_\mathcal{V},0,Y_\mathcal{V}/2,\Lambda_l)}{B(m_L,0,m_\mathcal{V},0,Y_\mathcal{V}/2)},\\
\hat{\Pi}_l^{(1)} &= \frac{A_R(m_\mathcal{T},0,Y_\mathcal{T}\sqrt{4/5},0,\Lambda_l)}{B(m_L,m_\mathcal{T},0,Y_\mathcal{T}\sqrt{4/5},0)},\\
\hat{\Pi}_\tau^{(1)} &= \frac{A_R(m_L,0, Y_\mathcal{T}\sqrt{4/5},0,\Lambda_\tau)}{B(m_L,m_\mathcal{T},0,Y_\mathcal{T}\sqrt{4/5},0)}\\
\hat{\Pi}_\nu^{(4)} &= \frac{A_R(m_L, 0, Y_\mathcal{V}/2,0,\Lambda_\nu)}{B(m_L, 0, m_\mathcal{V}, 0, Y_\mathcal{V}/2)}, &
\hat{\Pi}_\nu^{(6)} &= \frac{A_R(0,0,0,0,\Lambda_\nu)}{B(0,m_\mathcal{V},0,0,0)},\\
\hat{M}_\tau^{(1)} &= -\frac{A_M(m_L, m_\mathcal{T}, 0, Y_\mathcal{T}\sqrt{4/5},\Lambda_l, \Lambda_\tau)}{B(m_L, m_\mathcal{T}, 0, Y_\mathcal{T}\sqrt{4/5},0)},\\
\hat{M}_\nu^{(4)} &= -i\frac{A_M(m_L, m_\mathcal{V}, 0, Y_\mathcal{V}/2, \Lambda_l, \Lambda_\nu)}{B(m_L, 0, m_\mathcal{V}, 0, Y_\mathcal{V}/2)}
\end{aligned}
\end{equation}


\bibliographystyle{unsrt}
\bibliography{tex/refs}

\end{document}